\documentclass[preprint,onecolumn,12pt]{elsarticle}

\usepackage[utf8]{inputenc}
\usepackage{graphicx} 
\usepackage{epstopdf}
\usepackage{epsfig} 
\usepackage{amsmath} 
\usepackage{amssymb}  
\usepackage{xcolor}
\usepackage{placeins}
\usepackage{subfigure}
\usepackage{wasysym}
\usepackage{dsfont}

\begin{document}

\title{An empirical model for feedforward control of laser powder bed fusion}
\author[1]{Aleksandr Shkoruta\corref{cor1}%
}
\ead{ashkoruta@gmail.com}
\author[1]{Bumsoo Park}
\ead{park5@rpi.edu}
\author[1]{Sandipan Mishra}
\ead{mishrs2@rpi.edu}

\cortext[cor1]{Corresponding author}
\address[1]{Department of Mechanical, Aerospace, and Nuclear Engineering, Rensselaer Polytechnic Institute, Troy, NY 12180}

\maketitle

\section*{Highlights}
\begin{itemize}
    \item Laser scanning of narrow features in LPBF parts results in predictable changes in melt pool geometry as observed in coaxial melt pool images.
    \item An empirical model of this geometry-related behavior, validated on different geometries, is presented.
    \item A model-based feedforward controller is designed to correct for this behavior. The proposed controller reduces geometry-related deviations by a factor of two, as demonstrated experimentally.
\end{itemize}

\section*{Abstract}
While considerable progress has recently been made in real-time melt pool monitoring for laser powder bed fusion (LPBF), results in in-situ melt pool control are relatively sparse, a major reason being lack of suitable control-oriented models. This study demonstrates an empirical control-oriented model of geometry-dependent melt pool behavior, and subsequent melt pool regulation with a model-based feedforward controller for laser power. First, it shows that the melt pool ``footprint'' exponentially increases when the scan lines become shorter. The empirical model of this behavior is developed and validated on different geometries at different laser power levels. Second, the developed model is used to design a feedforward controller for obtaining optimal laser power profiles. This controller is then validated experimentally and is demonstrated to suppress the in-layer geometry-related melt pool signal deviations, for different part geometries.

\section{Introduction} \label{sec:intro}
Ensuring the quality of the laser powder bed fusion (LPBF) parts remains an open research problem, as the parts produced through this process are prone to defects such as cracks, porosity \cite{zhang2017defect}, and poor surface finish \cite{wang2013research,fox2016effect}. For the technology to become widely accepted and deliver on its promises, quality assurance must be guaranteed through process monitoring and control efforts. The subject of the melt pool control is of particular interest, as LPBF part quality is strongly related to the melt pool behavior \cite{King2015}.


Typically, the control problem in LPBF can be stated as a melt pool regulation problem: the goal is to compensate for unexpected deviations in the melt pool temperature or geometry, ultimately reducing effects such as overheating in acute corners, or dross formation. The regulator might be based on a process model or be model-free. 

Prior research has shown that the melt pool can be regulated with a feedback controller, based on a photodiode \cite{kruth2007feedback,Renken2019} or a camera \cite{Craeghs2011,Vasileska2020,Shkoruta2021} feedback signal. However, feedback control in LPBF is particularly challenging due to the high demands on the controller response time. Thus, the majority of literature focuses on the investigation of feedforward control strategies, such as a layer-to-layer data-driven control \cite{Shkoruta2019a}, and adjustment of the laser power based on geometry- and residual heat-based heuristics \cite{Yeung2019,Yeung2020-2}. A purely data-driven model to predict and control \textit{melt pool size} was also suggested \cite{Yeung2020-1}. Another example of data-driven control-oriented melt pool modeling was presented in \cite{ren2021gaussian}. There, Gaussian process regression (GPR) was used to model the melt pool dynamics, and feedforward control was applied to control the melt pool in a simulated environment. However, purely data-driven models are not readily admitting control design efforts and are hard to interpret. Lack of control-oriented models is well documented in LPBF, as the existing LPBF models are computationally expensive \cite{King2015} and are poorly suited for the in-situ process control. A good example of an interpretable, control-oriented model of the behavior of the melt pool cross-section is presented in \cite{Wang2020-penn}, where a model-based feedforward control application was experimentally demonstrated on a single track scale.  

Evidently, there is a lack of (a) control-oriented models and more specifically a lack of (b) explicitly geometry-aware models reported for LPBF. Geometric features of the scan layer, i.e. sharp corners or narrow areas, where part overheating might occur, can affect the melt pool behavior, leading to undesirable quality outcomes. Currently, such issues are mainly addressed through apriori process parameter optimization. It is thus of interest to model the effects of part geometry on the melt pool behavior. Therefore, this work focuses on the development of a geometry-aware control-oriented model of the melt pool behavior and on the experimental implementation of a geometry-aware model-based feedforward control for LPBF. 

The primary contributions of this work are the following: 
\begin{enumerate}
    \item Geometry-determined deviations of melt pool signatures, related to the decreasing scan line lengths, are reported, as observed in coaxial images.
    \item A control-oriented process model of these deviations is identified from empirical data, and experimentally validated. This model can be easily interpreted, is applicable to different scanning geometries and scanning patterns, and allows for in-layer model-based laser power control in a straightforward manner.
    \item A model-based feedforward controller built on the identified model is designed and validated. The controller output is found by solving an optimization problem formulated on a line-by-line scanning basis. The experimental application of the controller on a part scale decreased geometric deviations in melt pool signal by 50\%, for different part geometries.
\end{enumerate}
  
\section{Methods} \label{sec:methods}
\subsection{Hardware setup}  \label{sec:hw}

This research was performed on an open-architecture LPBF machine described in \cite{Shkoruta2019a}. The machine is equipped with a SCANLAB intelli\textit{SCAN}$_{de}$ 20 galvoscanner and a 400 $W$ NdYAG laser, and can build parts up to 50 $\times$ 50 $mm^2$ in cross-section from commercially available metal powders, e.g. stainless steel. The supervisory control of the machine is achieved via America Makes software \cite{AM_controller} augmented with in-house developed C++ code, while the low-level control of the scanning process, i.e. laser positioning and firing, is handled by the scanner control board. The scanning instructions for a layer are formatted as a text file containing a list of straight lines, each line defined by the start and end points, a laser power level, and a scanning speed value. Thus, arbitrary laser powers can be assigned to the layer scan on the line-by-line basis by preparing these files in advance. 



\begin{figure}[tb]
    \centering
    \includegraphics[width=0.5\linewidth]{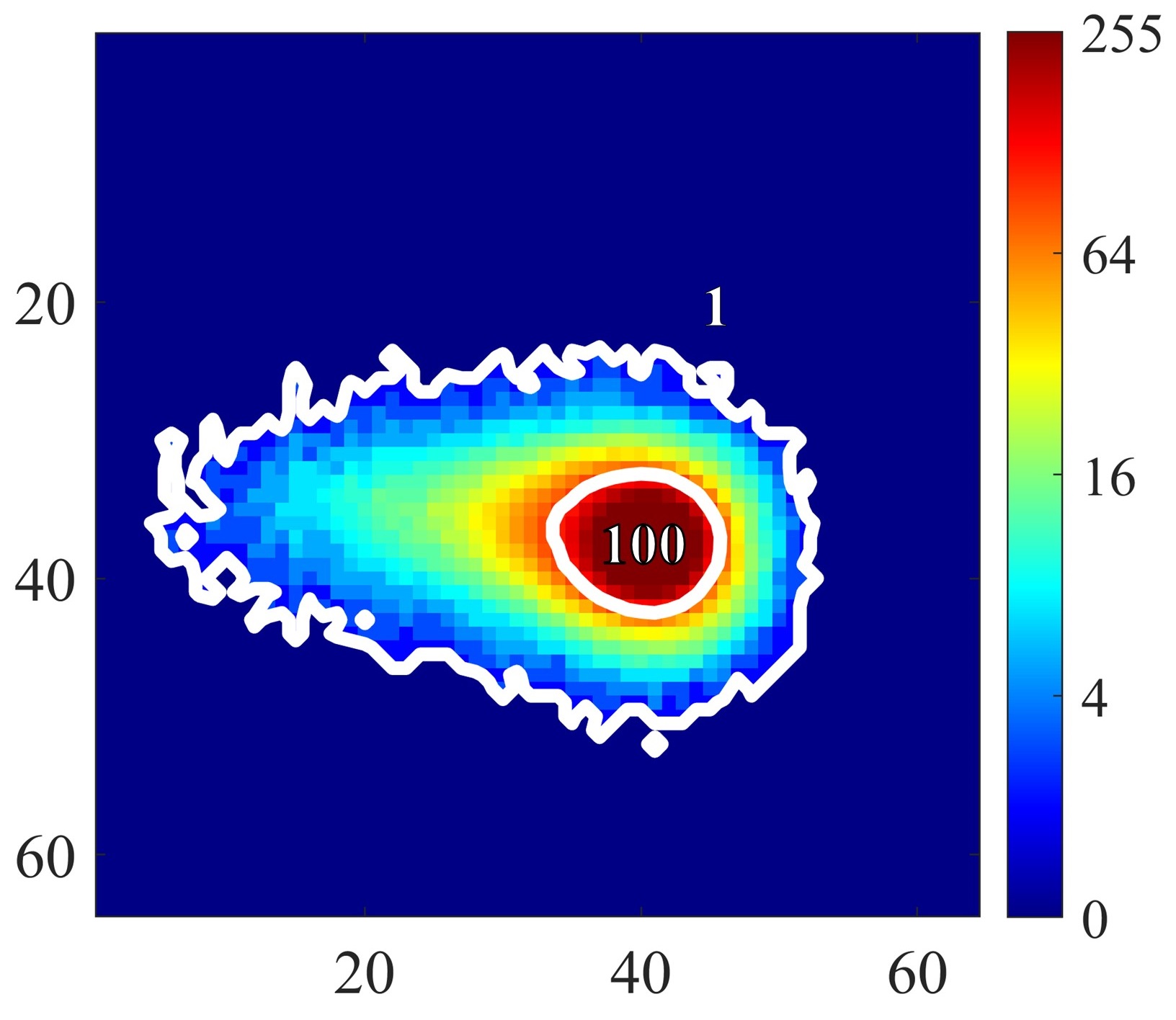}
    \caption{Typical melt pool image (false color, logarithmic scale of intensity). Level sets at $\alpha$ = 1 and $\alpha = 100$ are highlighted in white.}
    \label{fig:meltpool}
\end{figure}

To monitor the melt pool during the LPBF process, a coaxial camera-based setup, similar to described in literature \cite{kruth2007feedback,Grantham2016,Demir2018}, is integrated with the LPBF testbed. The Basler acA2000-165umNIR camera acquires 8-bit intensity images in near-infrared band ($800 - 950 \; nm$) at 2 kHz by looking at the melt pool through the laser scanning optics. 
Each image is 64 $\times$ 64 pixels in size, with an instantaneous field of view of 22 $\mu m$ per pixel. A typical melt pool image is shown in Figure \ref{fig:meltpool}.

\subsection{Image data processing}  \label{sec:coax}

Coaxial images are inherently multi-dimensional and hard to use for real-time control. One way to reduce the image dimensionality is to study the dimensions of its ``features'', e.g. an area of large low-intensity ``droplet'' of a near-zero emission intensity, or a small but bright ``hot spot'' in the center, as shown in Figure \ref{fig:meltpool}.
Both the ``droplet'' and the ``hot spot'' can be quantified by the following signals:

\begin{equation}
    C_\alpha = \sum_{r,c} \mathds{1}[I(r,c) \geq \alpha], \; \alpha = \{1,100\}
\end{equation}

\noindent where 
$r,c$ are row and column pixel coordinates, 
$I(r,c)$ is an image intensity of the pixel at $(r,c)$,
$\alpha$ is a threshold, 
and $\mathds{1}[...]$ stands for the indicator function, e.g. it is equal to 1 when its argument is true, and is 0 otherwise. For the ``droplet'', or the total ``footprint'' of the melt pool emission, an area of the level set at low threshold $\alpha = 1$ is appropriate, and for the ``hot spot'', higher $\alpha = 100$ is suitable. 

Ideally, the threshold level would be chosen such that the melt pool geometry is captured. However, this assignment is not as straightforward in practice \cite{Pacher2019}. 
In this work, thresholds $\alpha = \{1,100\}$ are picked based on the empirical observations of the coaxial images, and their physical meaning is open for interpretation. Other thresholds were not considered in this work.

\begin{figure}[htb]
    \centering
    \includegraphics[width=0.75\linewidth]{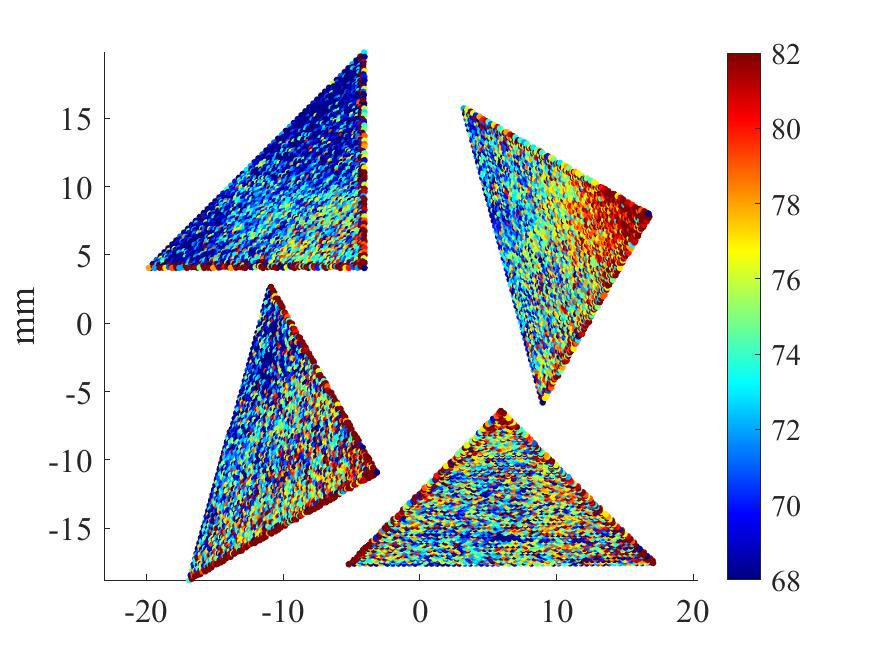}
    \caption{Example of a 1D coaxial signal (``hot spot'' area $C_{100}$, in pixels) as mapped to the nominal scanning position $(x,y)$ on the build plate.}
    \label{fig:xym}
\end{figure}

The melt pool location on the build plate at a given point in time was estimated based on the nominal scan pattern, assuming constant scanning velocity and the perfect trajectory tracking. Given the camera-assigned time stamp and the positional commands defining the layer scanning sequence, 1D melt pool signature such as $C_\alpha$ can be transformed from time series $C_{\alpha}(t)$ to a spatial map $C_{\alpha}(x,y)$, and plotted as a function of spatial coordinates in 2D, as illustrated in Figure \ref{fig:xym}. Such spatial maps are used in this work to illustrate the geometry-dependent melt pool behavior, and to demonstrate the effectiveness of the proposed feedforward control scheme.

\textit{Remark.} Note that at length scales below 500 $\mu m$, the mapping of the nominal scan pattern to a measurement is imprecise due to the lack of positional feedback. Thus, in the following analysis, if an image was found to belong to a line shorter than 0.5 $mm$, such image was excluded from the model identification.

\subsection{Test geometries}  \label{sec:parts}
\begin{figure}[htb]
    \centering
    \subfigure[]{    
    \includegraphics[width=0.47\linewidth]{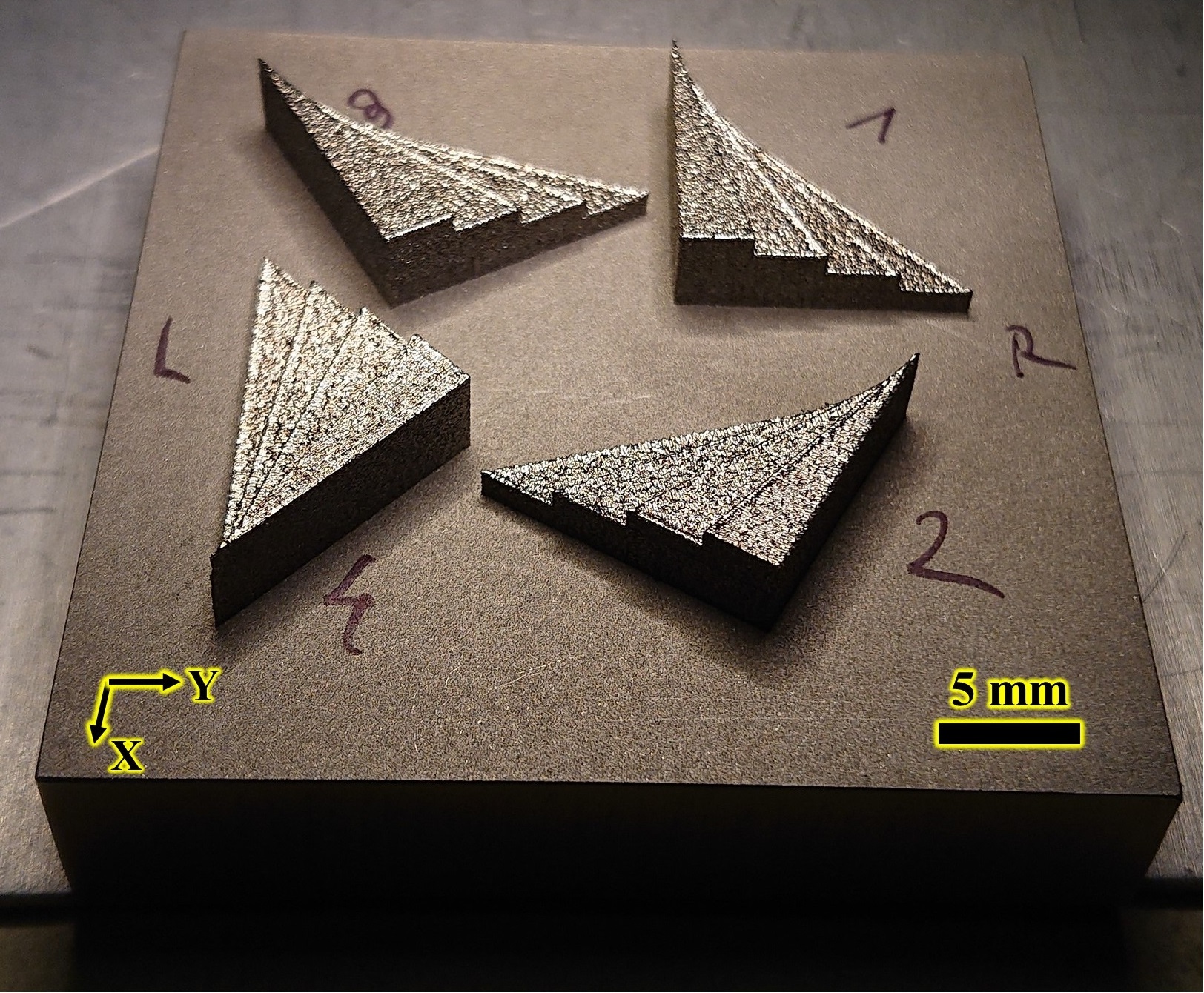}\label{fig:triangles}}
    \subfigure[]{
    \includegraphics[width=0.47\linewidth]{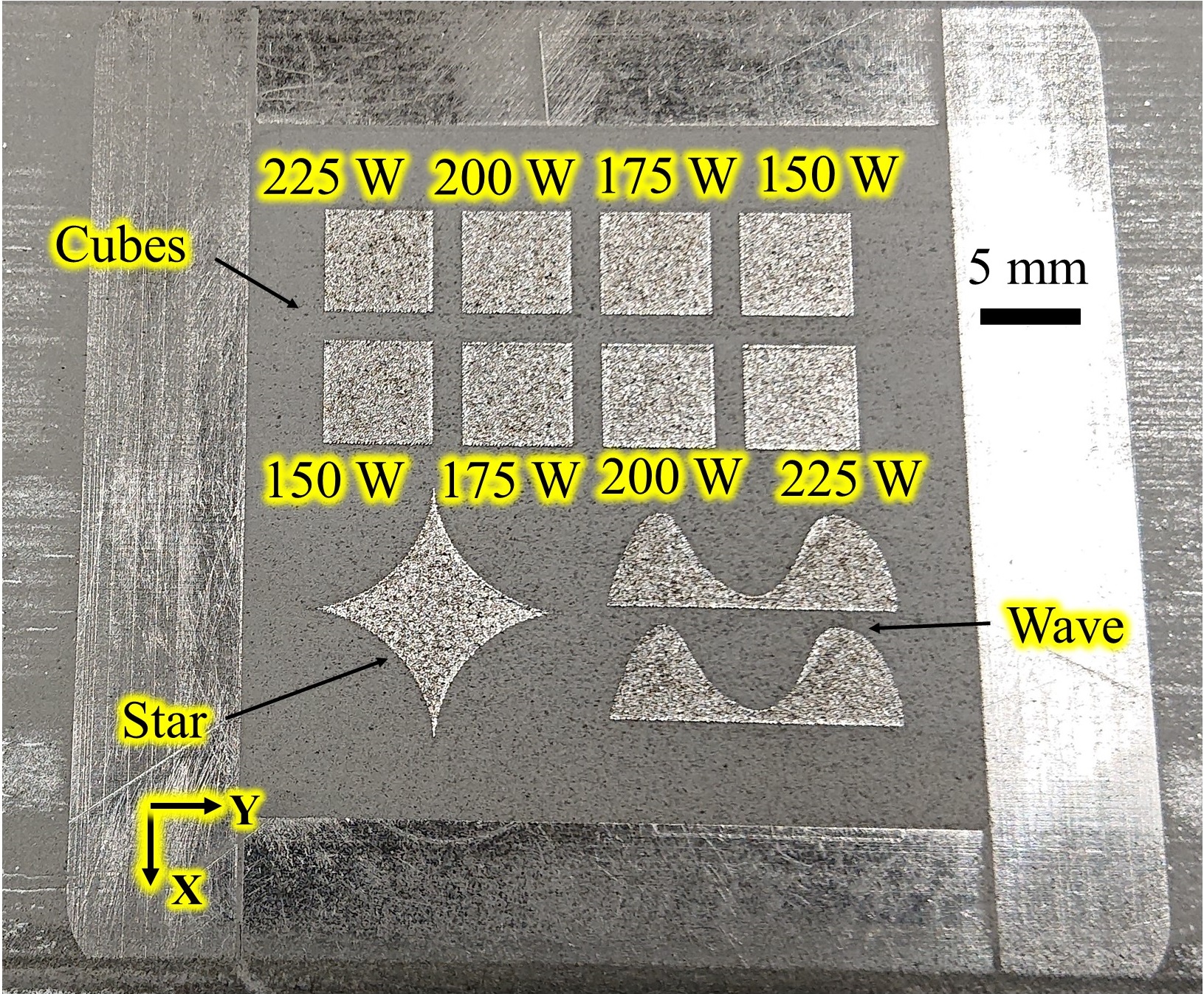}\label{fig:cubes}}
    \caption{ (a) Triangle prisms \textit{TP} test, as built on the LPBF testbed. (b) Parts built for the control-oriented model building and validation: \textit{Cubes}, \textit{Star}, and \textit{Wave} (with replication).}
\end{figure}

To build an empirical model of the melt pool behavior and to experimentally demonstrate model-based feedforward control, four kinds of geometries were employed in this work: stacked triangular prisms, cubes, a star, and a wave. Figure \ref{fig:triangles} shows the as-built triangular prisms, and Figure \ref{fig:cubes} illustrates the other geometries. All parts were produced from a cobalt chrome alloy. Detailed descriptions of the geometries and the process parameters are presented below.

\begin{figure}[hbt]
    \centering
    \includegraphics[width=0.75\linewidth]{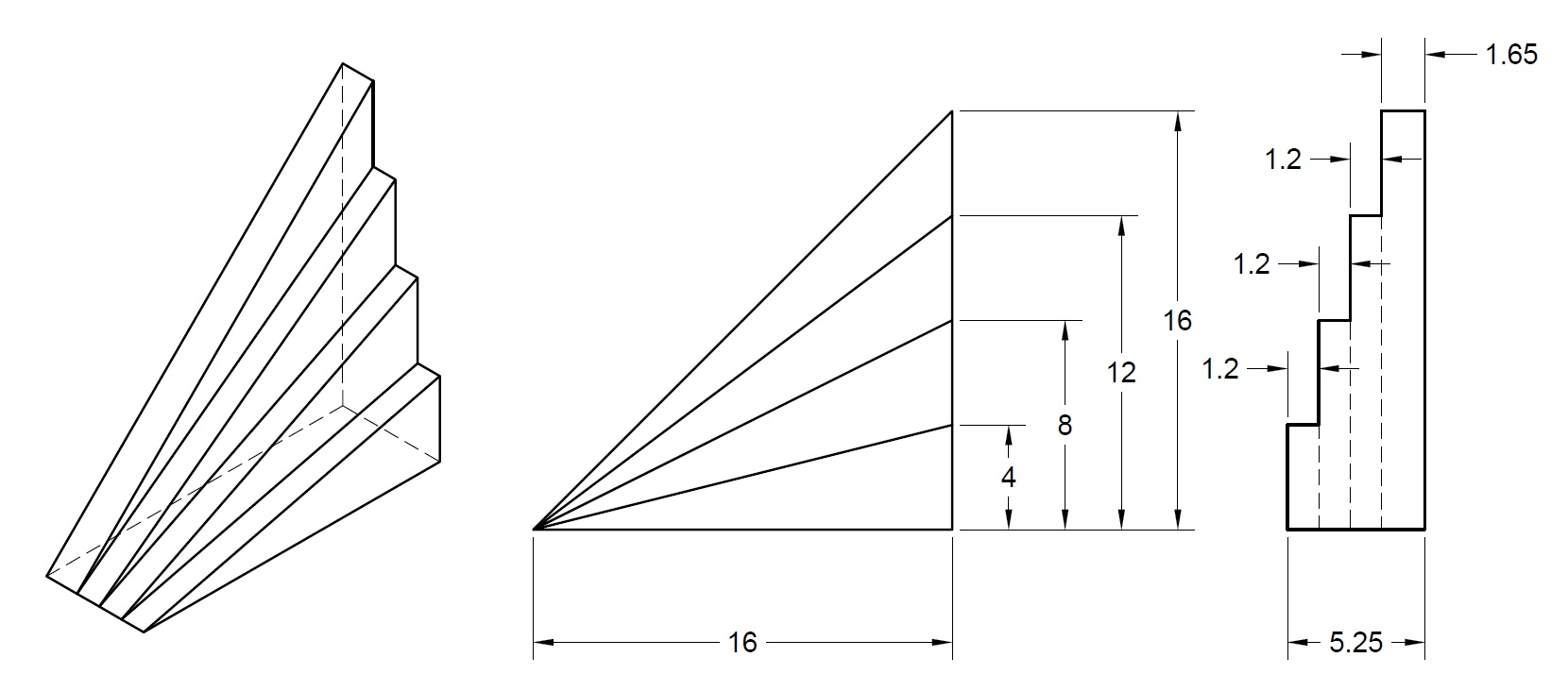}
    \caption{Triangle prism geometry for the  evaluation of the corner scanning effect. Dimensions in mm.}
    \label{fig:triangle_dwg}
\end{figure}

\begin{figure}[hbt]
    \centering
    \includegraphics[width=0.8\linewidth]{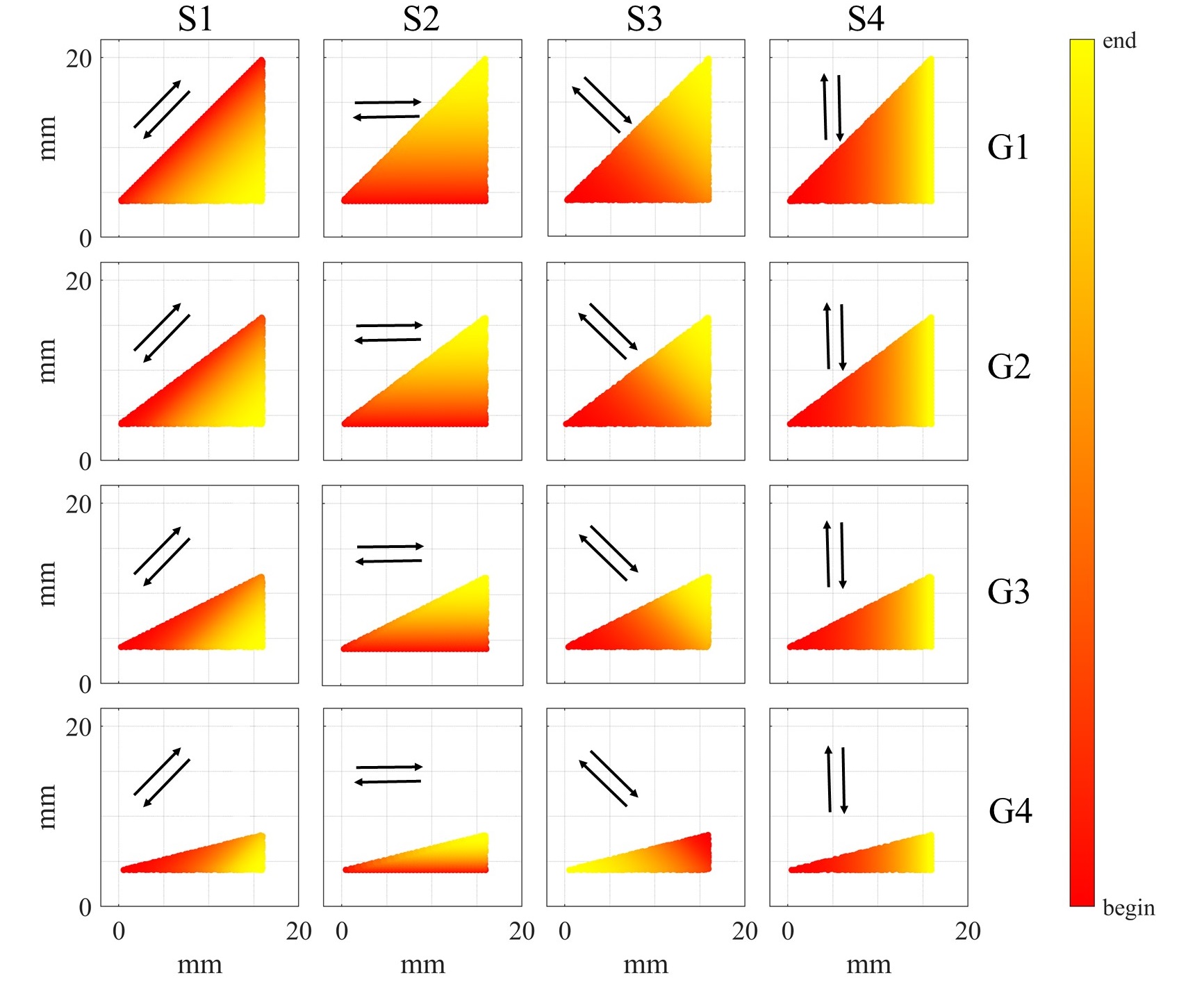}
    \caption{Scan patterns for all \textit{G}, \textit{S} in the triangular prism (\textit{TP}) test. Arrows indicate scanning direction. Color change indicates the scan progression from the scanning start to its end.}
    \label{fig:ilc_scan}
\end{figure}

\paragraph{Triangular prisms} A prismatic part in Figure \ref{fig:triangle_dwg} was designed to investigate if a quick return of the laser in proximity of the previously scanned track would influence melt pool behavior. The part consists of 4 stacked prisms, triangular in cross-section, denoted as \textit{G1-G4}. Each prism was scanned at 4 hatching angles: $0^{\circ}$, $45^{\circ}$, $90^{\circ}$, and $135^{\circ}$ with respect to the part's edges, denoted as \textit{S1-S4} correspondingly (Figure \ref{fig:ilc_scan}). 
The whole build included 4 parts (\textit{P1-P4}) of the same geometry, identical by all means other than the rotation with respect to the machine coordinate system. In-plane rotation was introduced to control for the directions of powder recoating and the gas blower cross-flow.

For all parts \textit{P}, prisms \textit{G}, and scans \textit{S} parameters were set as: laser power 225 $W$, scanning velocity 800 $mm/s$, hatch spacing 90 $\mu m$. Continuous meander (snake-like) scan pattern was used: laser was always on and no scan ``jumps'' were performed. In total, 174 layers were built for each part \textit{P1-P4}, 
with each scan pattern \textit{S1-S4} repeating at least 10 times. 

Parts \textit{P1-P4}, as built and still attached to the build plate, are shown in Figure \ref{fig:triangles}. This build will be referred to as \textit{TP} (triangular prism) test from now on.

\paragraph{Cubes of the varying laser powers} The \textit{TP} test was performed a single power level and did not provide enough data for the power-dependent model building. Therefore, cubes of varying laser powers were built to acquire enough data for a control-oriented process model.

Four different laser powers (150, 175, 200, 225 $W$) were used to build 8 cubes, with each power replicating twice. Each cube was scanned at 4 angles: $30^{\circ}$, $135^{\circ}$, $60^{\circ}$, and $135^{\circ}$ again. The hatch spacing and scanning velocity were repeated from the \textit{TP} test, and scanning was performed in the meander fashion without jumps, as well. Figure \ref{fig:cubes} illustrates the cubes with their laser powers indicated. This test is referred to as \textit{Cubes} test in the following text.

\paragraph{Star and wave} Lastly, two additional geometries were used for in-situ demonstration of the feedforward control algorithm. No data from these parts was used for the model building at any point.

\begin{figure}[htb]
    \centering
    \subfigure[]{
    \includegraphics[width=0.45\linewidth]{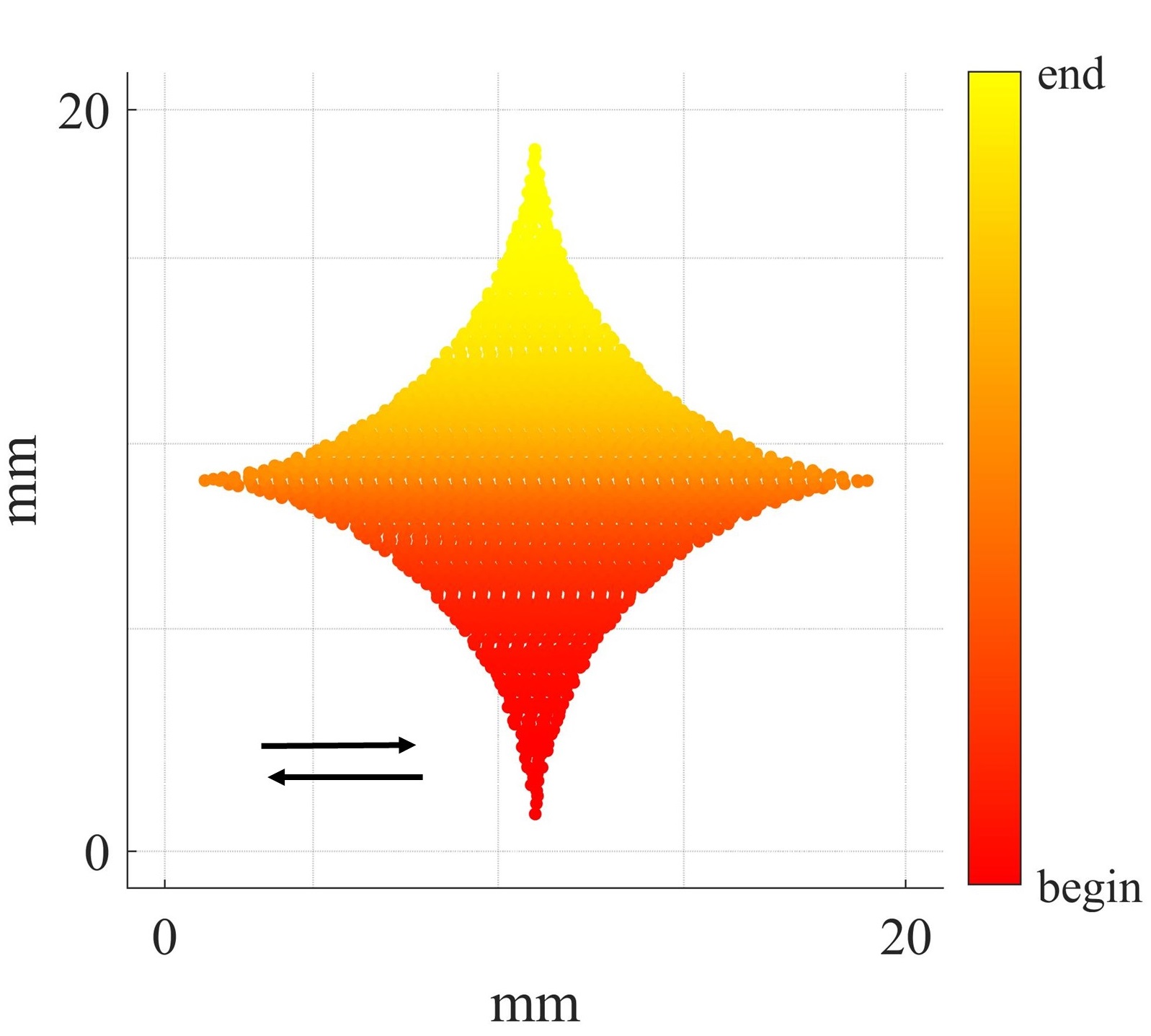}}
    \subfigure[]{
    \includegraphics[width=0.45\linewidth]{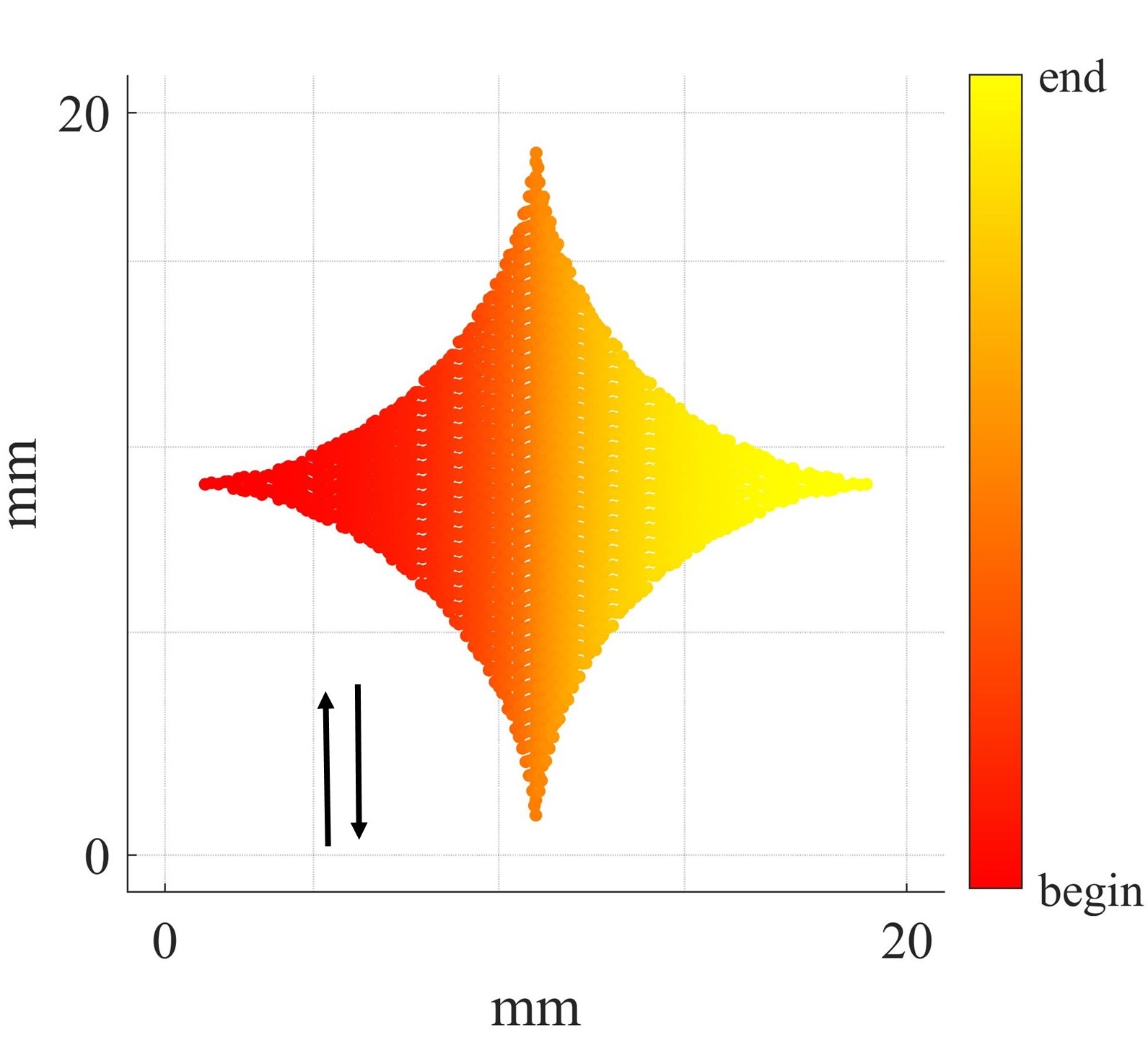}}
    \caption{Two scan patterns for \textit{Star} part: (a) horizontal and (b) vertical. Both scans proceed continuously from one corner to the opposite one.}
    \label{fig:star_scan}
\end{figure}

\begin{figure}[htb]
    \centering
    \includegraphics[width=0.75\linewidth]{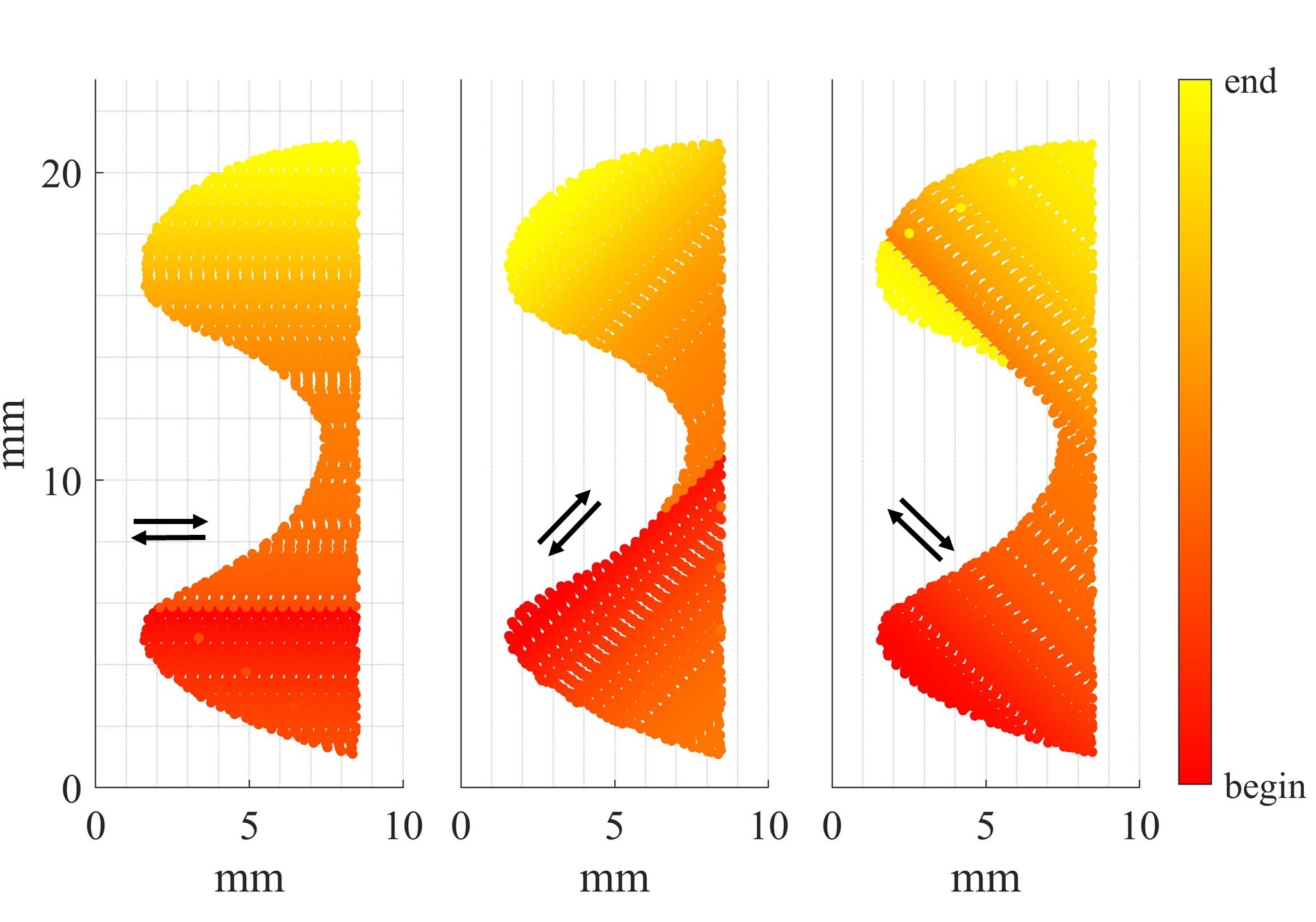}
    \caption{Three unique scan patterns for \textit{Wave} part. Note that scans are discontinuous: laser re-positions across the part in some cases.}
    \label{fig:wave_scan}
\end{figure}

Geometries in \textit{TP} and \textit{Cubes} tests have straight boundaries and linear scan patterns. Thus, two validation geometries were designed with curved edges: one part (\textit{Star}) was designed to have sharp corners, and the other (\textit{Wave}) was designed to have a narrow ``necking'' section. \textit{Star} was scanned at $0^{\circ}$ and $90^{\circ}$ (horizontally and vertically), while \textit{Wave} was scanned at 3 different angles (cycling through $0^{\circ}$, $45^{\circ}$, $0^{\circ}$, $135^{\circ}$), as illustrated in Figures \ref{fig:star_scan} and \ref{fig:wave_scan}. Note that while \textit{Star} was scanned continuously, \textit{Wave} was scanned in separate sections and laser was re-positioning (jumping) during scans, as decided by the internal algorithm of the slicer. Nominal laser power was set to 200 $W$, while the scanning speed and hatch spacing remained the same as in other tests.


\subsection{Model identification and validation}
Data acquired from the \textit{TP} test was used for exploratory data analysis. Based on that, an exponential model structure to predict $C_1$ was selected: 

$$C_1 = a + b\exp(-l/c)$$
\noindent where $a,b,c$ are the fit parameters, and $l$ is the regressor - the scan line length. 

The coefficients of the exponential model were parameterized with laser power based on the data acquired from the \textit{Cubes} test. The first 30 layers of the \textit{Cubes} were used as training data: they were split into 6 chunks of 5 layers each, and the exponential fit parameters were identified on each of those 6 chunks. 
As such, there were $2\times6 = 12$ sets of identified parameters for each laser power, and therefore $12\times4 = 48$ data points to produce a nominal model with linear regression analysis. 

The parameterized model was subsequently validated on the last 3 layers (24 separate scans) of the \textit{Cubes} which were not used for the model identification. Due to the high-frequency variations inherent to the $C_1$ signature, validation data was smoothed, such that the fit to a geometry-related \textit{trend} was found: data was filtered with a median filter with a window length of 150 samples, approximately $1/10$ of a layer scan.
\FloatBarrier

\section{Results}
The results of this research can be summarized as follows:
\begin{enumerate}
    \item A meandering scan path results in a repeatable geometry-determined behavior of the coaxial image ``footprint'' $C_1$, where ``footprint'' grows exponentially with decreasing scan line length;
    \item This behavior can be captured by a dynamic model
$$C_1(n) = C_\infty(p_n) + \Delta C(p_{n-1})e^{-\frac{l_{n-1}}{r(p_{n-1})}}$$ 
        
        where $n$ is the scan line index, $l$ is the scan line length, $p$ is the laser power assigned on the line-by-line basis, and $C_\infty$, $\Delta C$, and $r$ are well-defined polynomial functions of laser power;
    \item Based on the model above, optimization-based feedforward laser power controller reduced geometry-induced variations in melt pool ``footprint'' by 50\%.
\end{enumerate}
\noindent The following Sections will present these results in greater detail.

\subsection{Exponential corner-related behavior}
The coaxial images acquired from the scanning of triangular geometries (the \textit{TP} test, as described in Section \ref{sec:methods}), repeatably showed geometry-dependent variations. The specific behavior depended on the coaxial signature selected, as the ``footprint'' $C_1$ and ``hot spot'' $C_{100}$ behaved differently: the ``footprint'' area would reliably increase in corners, while $C_{100}$ would not. However, $C_{100}$ would repeatably increase at the edges of the triangles. While the $C_{100}$ behavior appeared to be highly localized, $C_1$ showed in-layer dynamics and thus became the focus of this research. Only $C_1$ signal modeling and regulation is considered in this section. While the regulation of $C_{100}$ is also of interest, it is not considered here and should be the subject of future work.

\begin{figure}
    \centering
    \subfigure{%
    \includegraphics[width=0.49\linewidth]{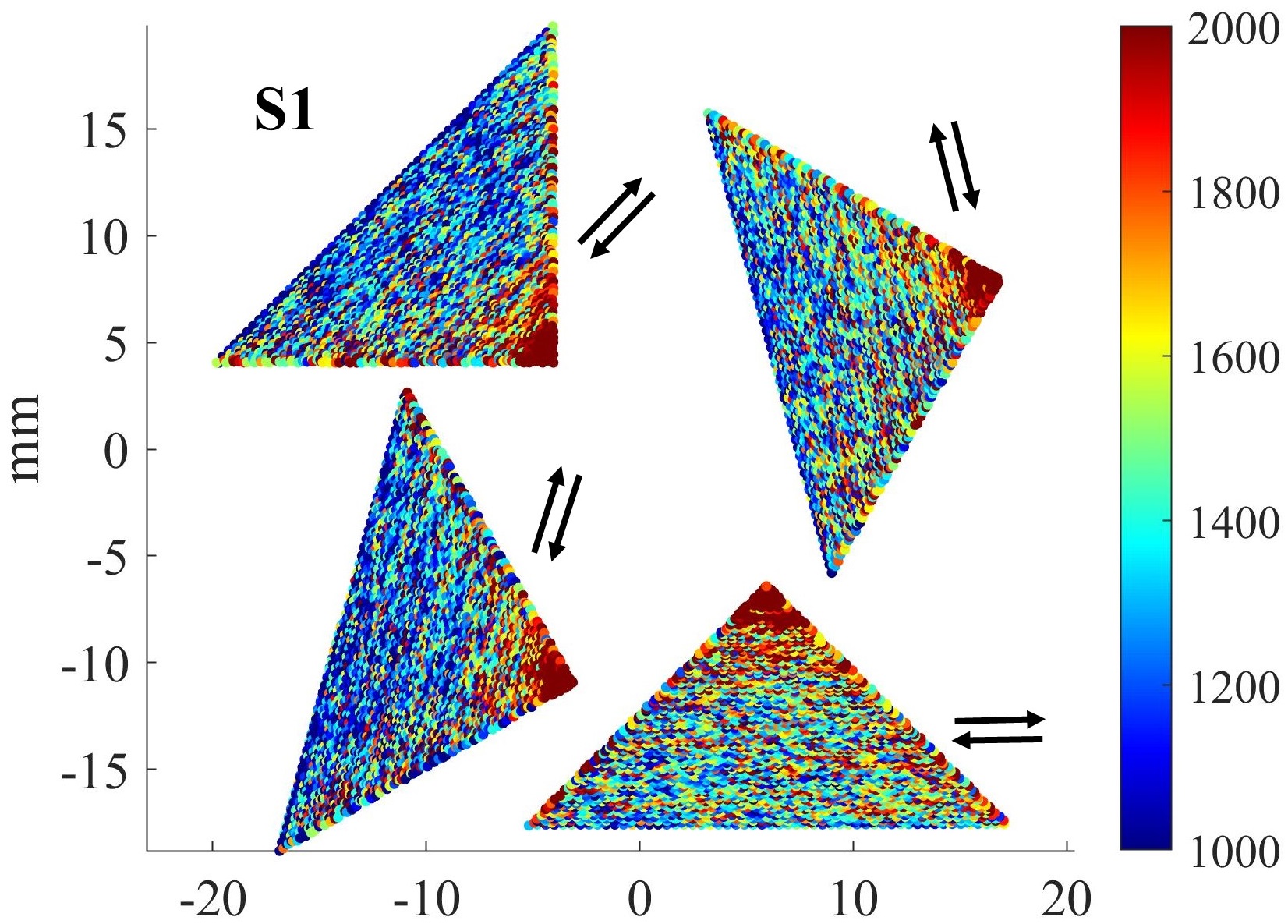}}
    \subfigure{%
    \includegraphics[width=0.49\linewidth]{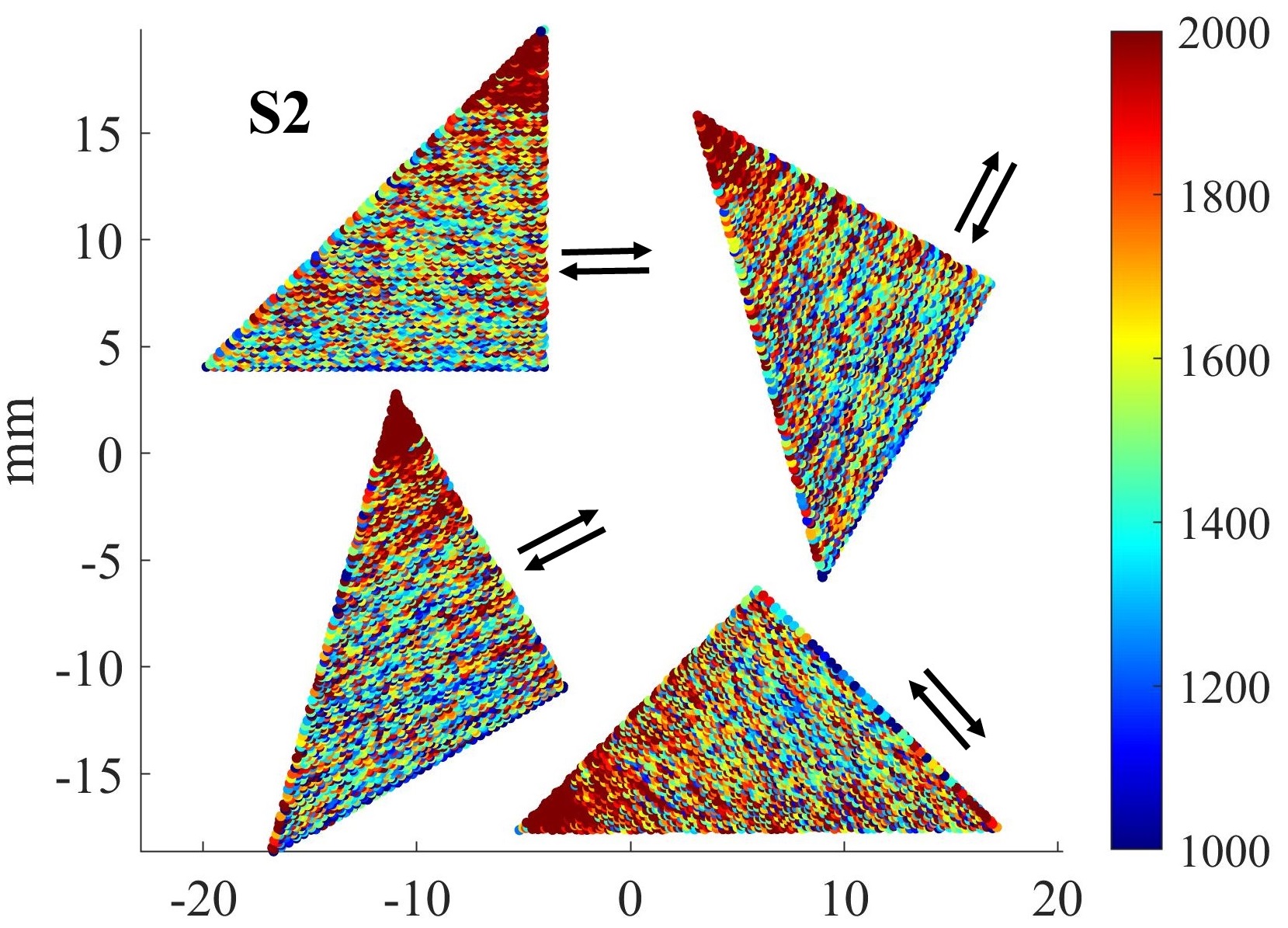}}
    \hfill
    \subfigure{%
    \includegraphics[width=0.49\linewidth]{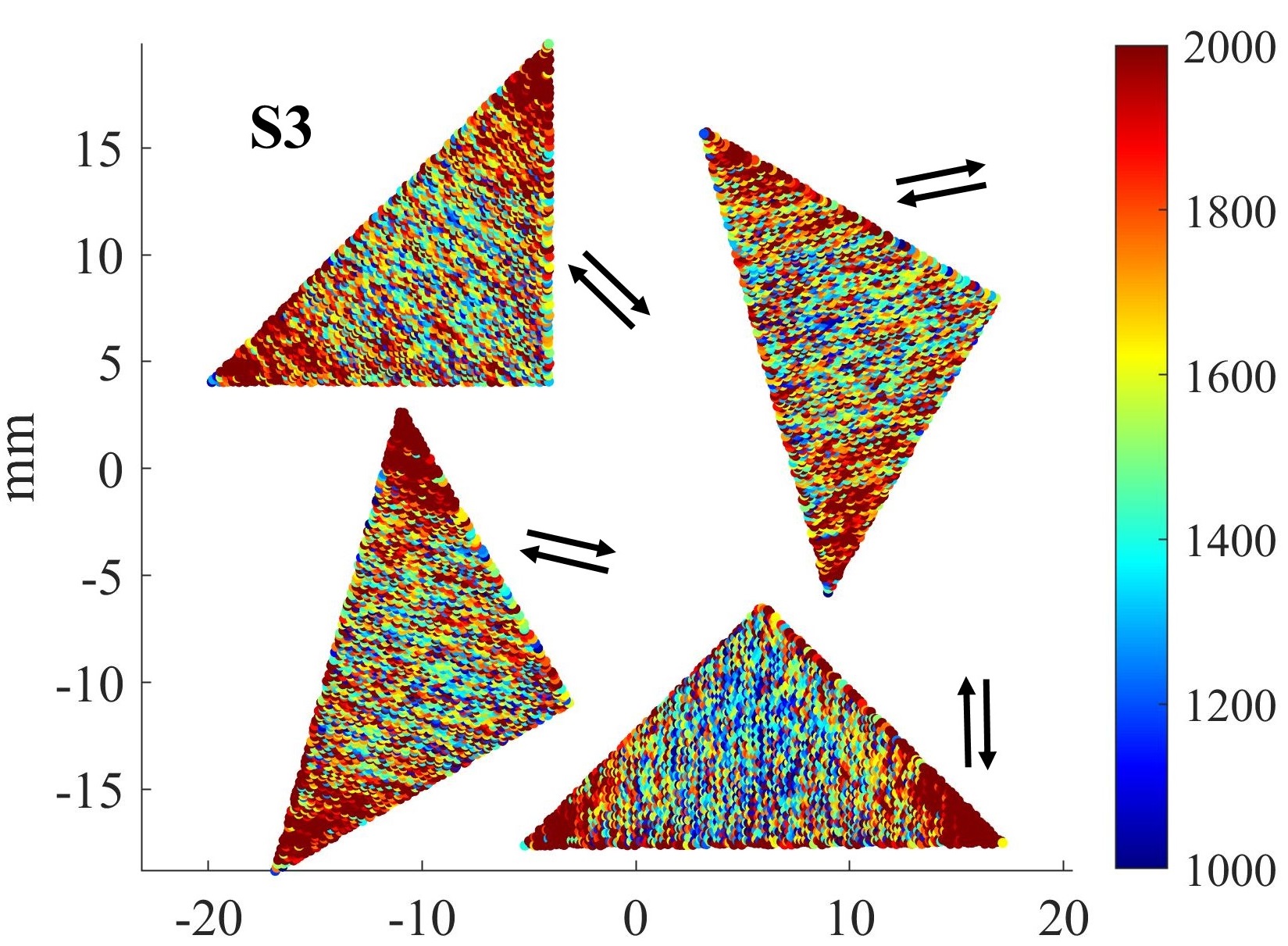}}
    \subfigure{%
    \includegraphics[width=0.49\linewidth]{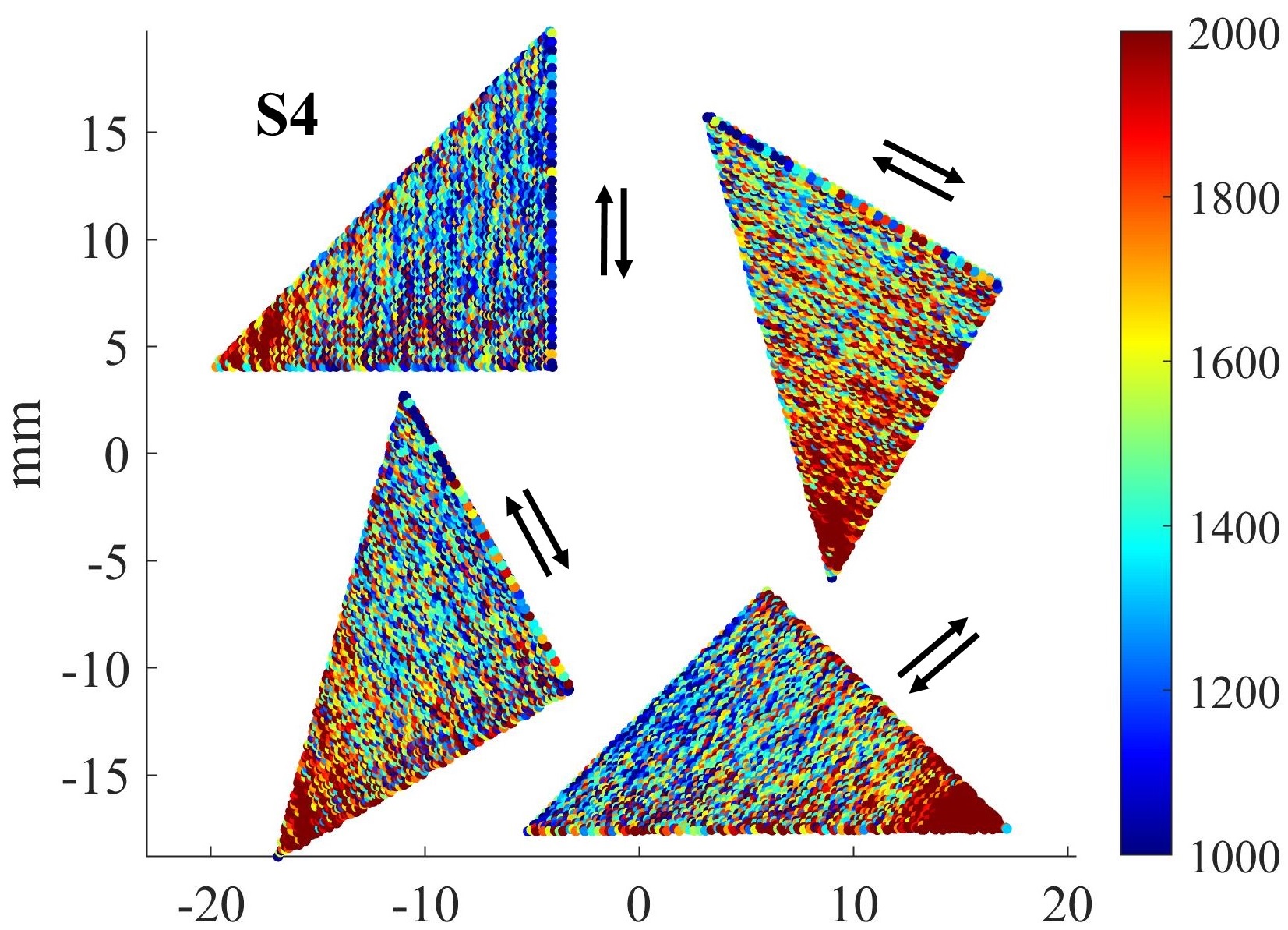}}
    \caption{Corners appear to cause the increased levels of $C_1$ coaxial signature. Scanning direction is indicated by the black arrows for each of 4 parts. The signal increase appears to be independent of the laser travel direction and only depends on the scanning direction relative to the part features. A representative selection of 4 different layers is shown.}
    \label{fig:corners}
\end{figure}

The $C_1$ signal for 4 different scan patterns \textit{S1-S4} is plotted in Figure \ref{fig:corners}, where the scanning direction for each triangle is indicated by the black arrows. Clearly, $C_1$ increased in the corners of all 4 triangles, and this effect is invariant of the triangle orientation and the triangle size. For the sake of brevity, only the largest triangles (\textit{G1}) are shown in Figure \ref{fig:corners}.

%

For each scan direction \textit{S1-S4}, the coaxial signature $C_1$ increased in part corners in a qualitatively similar manner, for all parts \textit{P1-P4} and all prisms \textit{G1-G4}. It appeared that this effect was related to the local geometry and thermal history, i.e. local accumulation of heat. Thus, to predict this observed behavior of $C_1$, different regressors, such as total time of scanning, coordinate- and time-based features, and a history of $C_1$ prior to the current measurement were tried. The $C_1$ signature was plotted against different features in different ways, and machine learning approaches were also tried. In the end, it appears that $C_1$ evolves exponentially depending on the scan line length, for all \textit{P}, \textit{G}, \textit{S}. 

\begin{figure}[htb]
    \centering
    \subfigure[]{%
    \includegraphics[width=0.48\linewidth]{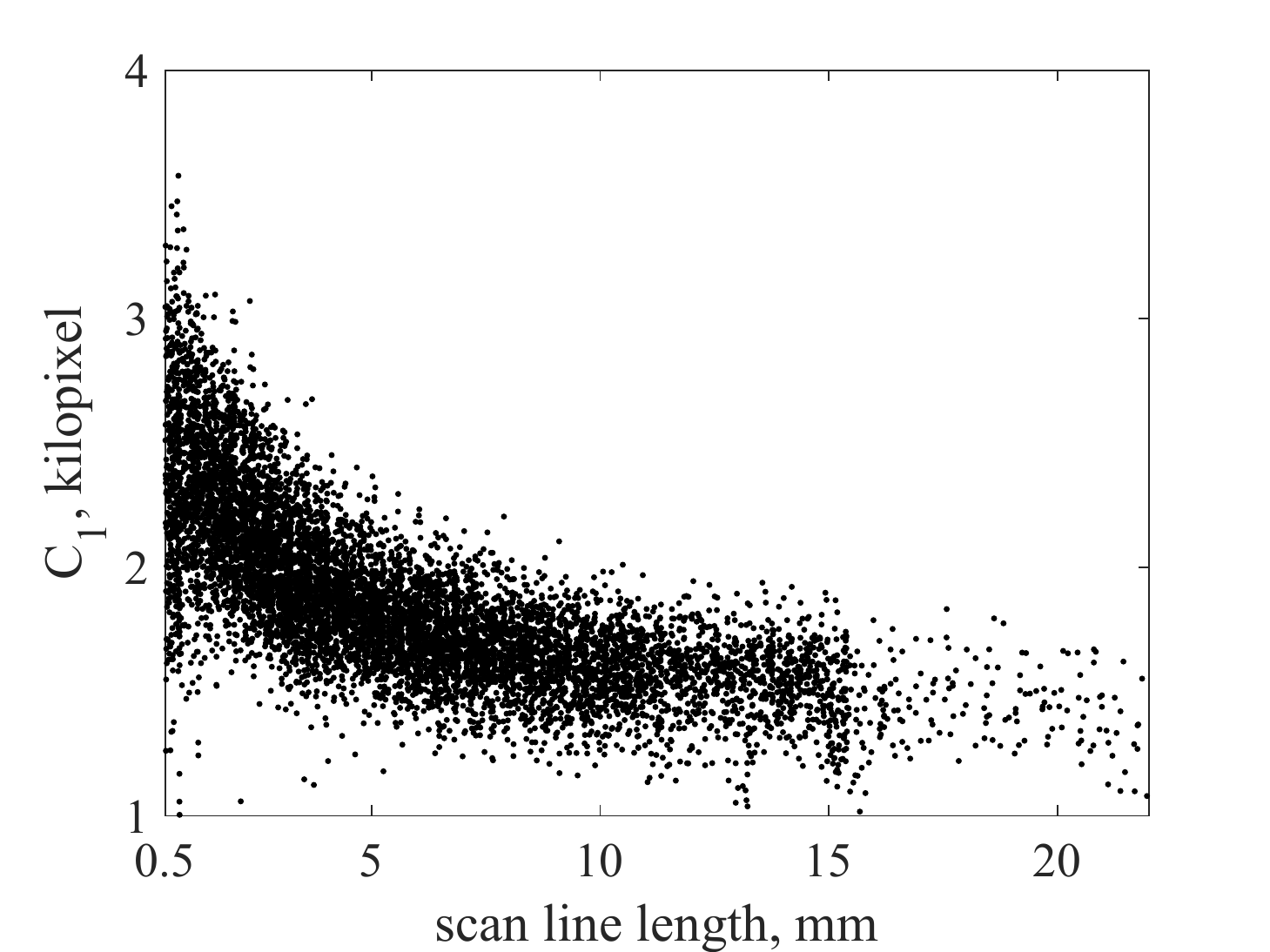}\label{fig:expon_trend}}
    \subfigure[]{%
    \includegraphics[width=0.48\linewidth]{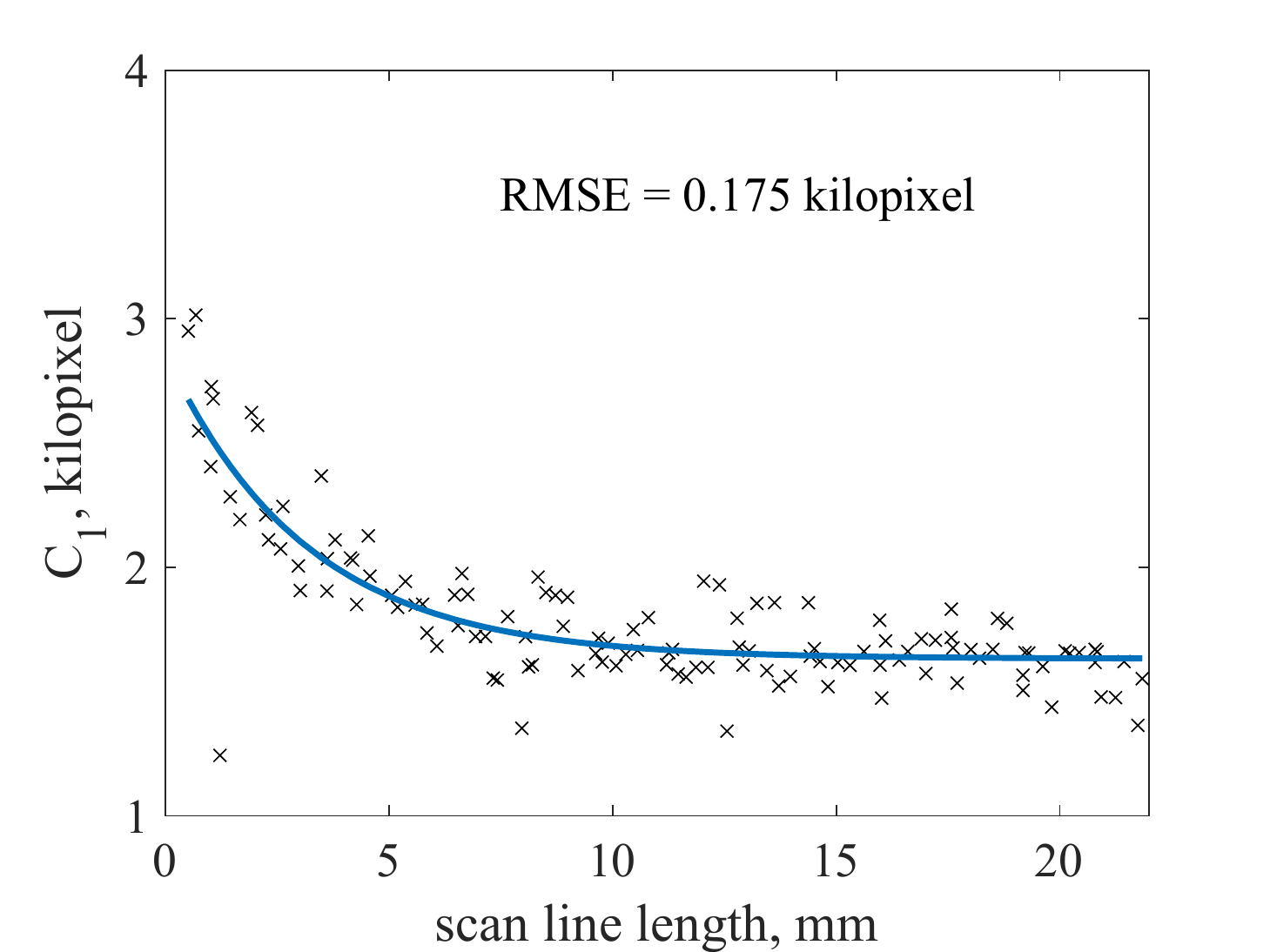}
    \label{fig:expon_fit}}
    \caption{(a) Measurements of $C_1$ from different \textit{P}, \textit{G}, \textit{S} combinations plotted against scan line length. A clear exponential trend is observed. (b) Exponential model (\ref{eq:exp_structure}) as fitted to one scan \textit{P4},\textit{G1},\textit{S1}.}
\end{figure}

Figure \ref{fig:expon_trend} shows the scatter plot of the experimental data to support the claim that the relationship is exponential. Here, one layer from each \textit{P}, \textit{G}, \textit{S} (64 separate layers) is presented. For each coaxial image, $C_1$ was evaluated, and the corresponding scan line index was estimated, as described in Section \ref{sec:methods}. Despite coming from significantly different scan patterns \textit{G1-G4} and different scanning directions \textit{S1-S4}, all points still fall on the same curve. This curve approaches a defined lower limit, as the scan length increases, and rapidly grows as the scan line length decreases.

Examining Figure \ref{fig:expon_trend}, a suitable model structure is
\begin{equation}
    C_1(l) = C_\infty + \Delta C \exp(-l/r)
    \label{eq:exp_structure}
\end{equation}
\noindent where $l$ is the scan line length, and $C_\infty$, $\Delta C$, and $r$ are the fit parameters. Fitting this exponential model to one particular scan \textit{P4,G1,S1} results in a root mean square error (RMSE) value of 175, and $R^2$ value of 0.68, as shown in Figure \ref{fig:expon_fit}. This model then performs similarly well on other parts \textit{P}, prisms \textit{G}, and scan directions \textit{S}. Because the exponential model is further extended in the following sections, details of the validation of the model (\ref{eq:exp_structure}) are omitted for brevity.

\subsection{Exponential model parameterized by laser power} \label{results:model}
The systematic variation in $C_1$ (near the corners, for example) can be addressed by changing the input laser power, such that $C_1$ is regulated to a constant value across the whole layer, regardless of the scan line length. However, the model (\ref{eq:exp_structure}) is only valid for a \textit{particular} power level and should somehow incorporate variable laser power to be suitable for process control. To extend the exponential model (\ref{eq:exp_structure}), measurements acquired during the \textit{Cubes} test were used.

\begin{figure}[hbt]
    \centering
    \includegraphics[width=0.75\linewidth]{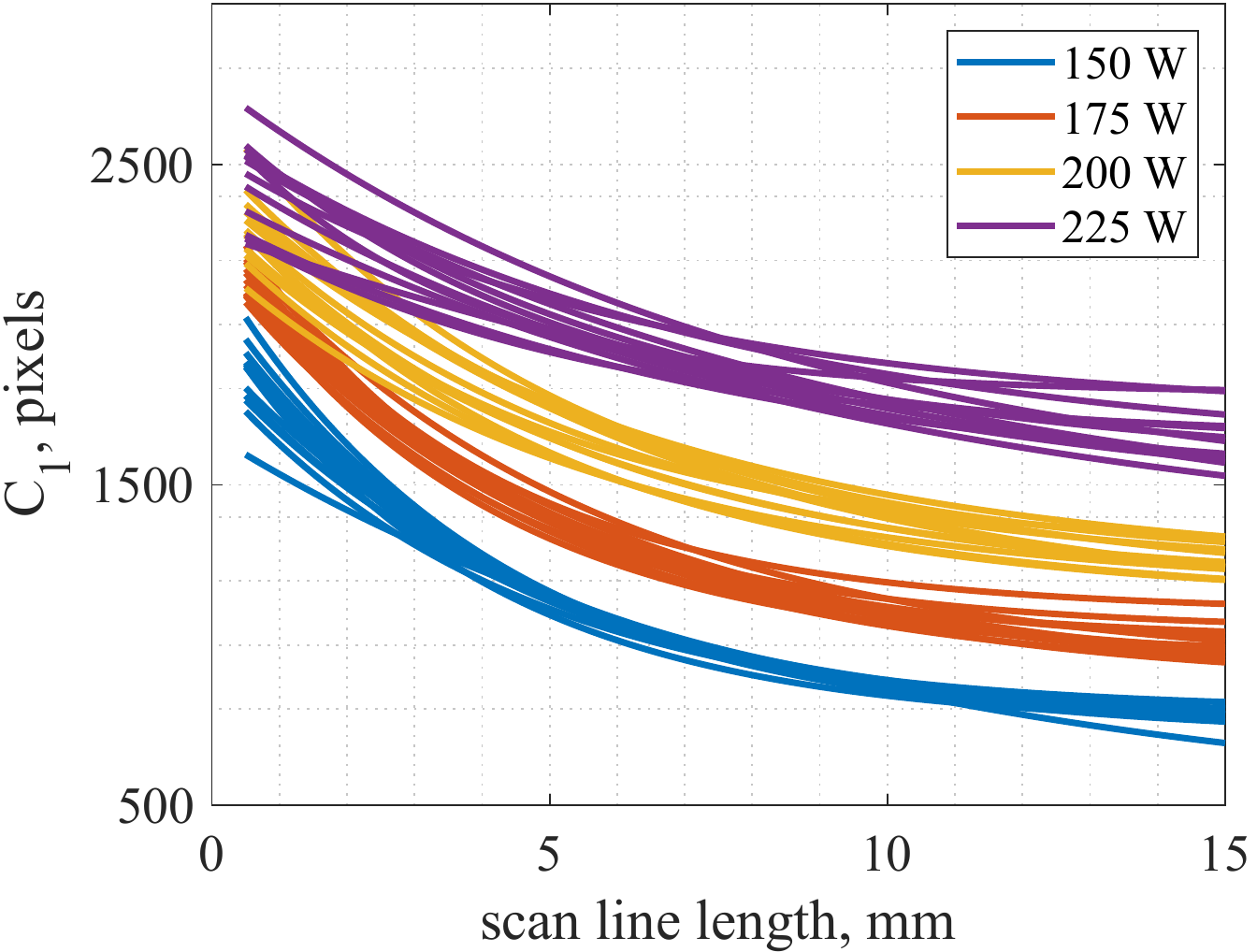}
    \caption{The dependence of the exponential fit on laser power.}
    \label{fig:expon_fit_vs_power}
\end{figure}

The \textit{Cubes} test demonstrated that corner-related exponential behavior also occurs at different laser powers. Figure \ref{fig:expon_fit_vs_power} shows all of the identified exponential fits for $C_1(l)$ as produced for different power levels (with replication, as explained in Section \ref{sec:methods}). Evidently, coefficients of exponential fit can be parameterized as functions of laser power: $C_\infty = C_\infty(p)$, $\Delta C = \Delta C(p)$, $r = r(p)$.

\begin{figure}[hbt]
    \centering
    \includegraphics[width=0.61\linewidth]{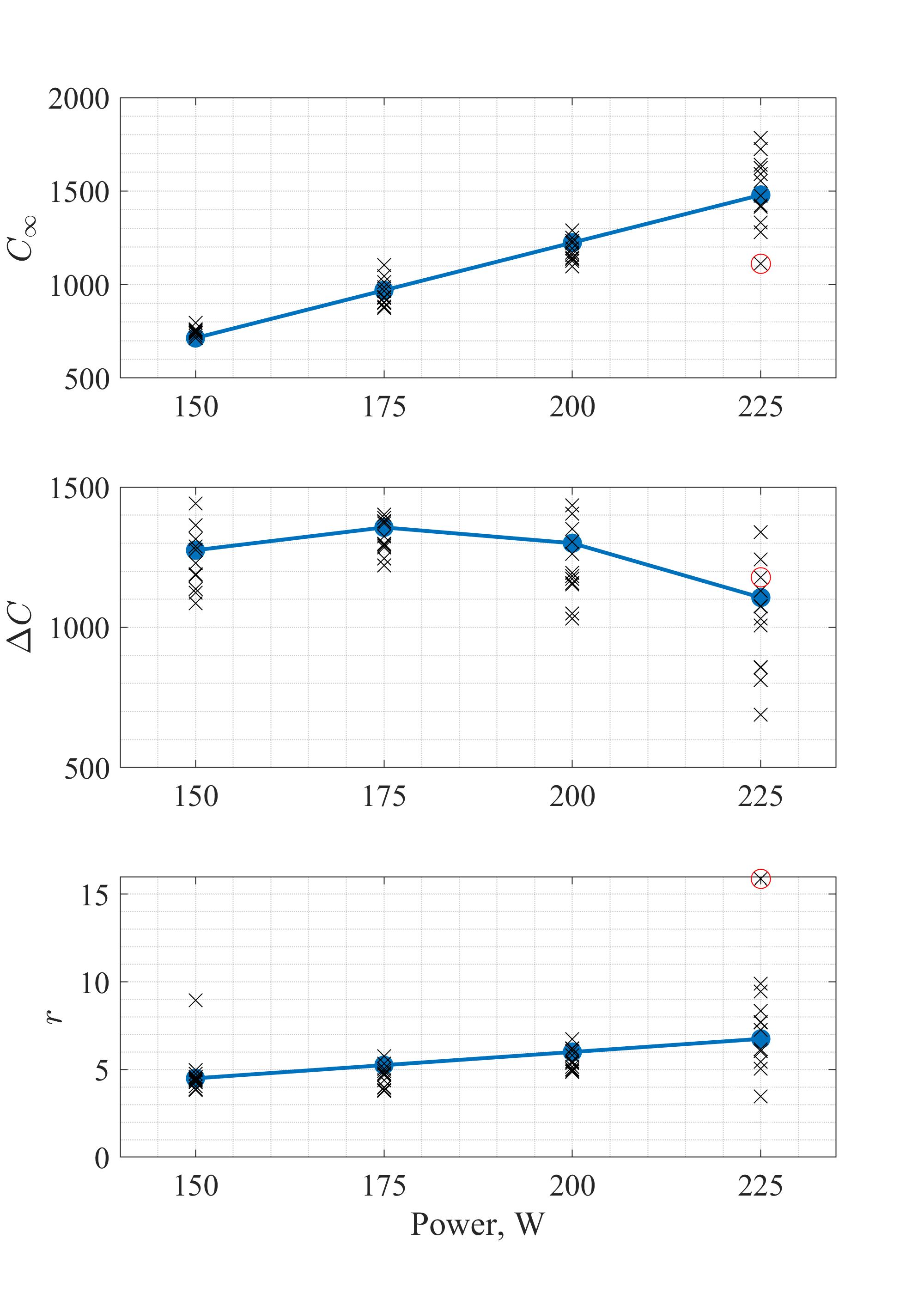}
    \caption{Fit parameters $C_\infty$, $\Delta C$, $r$ as functions of laser power. Parameter values, as estimated from data, are shown as crosses. The red circle indicates an outlier excluded from the fit. Shaded regions indicate the confidence band at 0.05 level.} 
    \label{fig:coeff_regr}
\end{figure}

Figure \ref{fig:coeff_regr} shows how the identified model parameters depend on the laser power level. A linear regression was used to model the parameter dependency, and one outlier was dropped out of data based on Cook distance analysis. Notably, there is a larger spread in identified parameters at the largest laser power of 225 \textit{W}. The cause of this spread is unknown and should be investigated further in future work.

$C_\infty$ was found to be linear with laser power ($R^2 = 0.895$). The estimated slope is $10.6 \pm 0.5$\footnote{Standard error is used whenever uncertainty estimate is given.}, estimated intercept is $-880 \pm 102$. Meanwhile, estimates for $\Delta C$ and $r$ had larger spreads. The decay rate parameter $r$ appeared to weakly depend on power but linear trend fits data better than a constant (median) value, thus $r$ was modeled as a linear function ($R^2 = 0.307$). Estimated slope is $0.030 \pm 0.007$, and intercept was estimated to be $-0.14 \pm 1.3$ (with \textit{p}-value of 0.9). Due to the low confidence in the intercept estimate and its relatively low contribution in the studied power range\footnote{$r \geq 4.5$ at laser powers above 150 $W$, thus intercept value of $-0.14$ has a relatively small contribution to the value of $r$.}, intercept term was disregarded. Finally, for the range parameter $\Delta C$, a downward curve to the data was observed. $R^2 = 0.456$ in the quadratic model, which is better than a linear fit, and residuals appeared to be normally distributed. Fit coefficients were found to be: intercept $-2268 \pm 1031$, linear term $41 \pm 11$, and quadratic term $-0.12 \pm 0.03$. Therefore, the following fits were obtained for the exponential model parameters:

\begin{equation*}
    C_\infty(p) = 10.6p - 880
\end{equation*}
\begin{equation}
    \Delta C(p) = -0.12p^2 + 41p - 2268
\end{equation}
\begin{equation*}
r(p) = 0.03p    
\end{equation*}

The resultant model is thus 

\begin{equation}
C_1 = C_\infty(p) + \Delta C(p)\exp(-l/r(p))    
\label{eq:model}
\end{equation}

\begin{figure}[t!]
    \centering
    \includegraphics[width=0.5\linewidth]{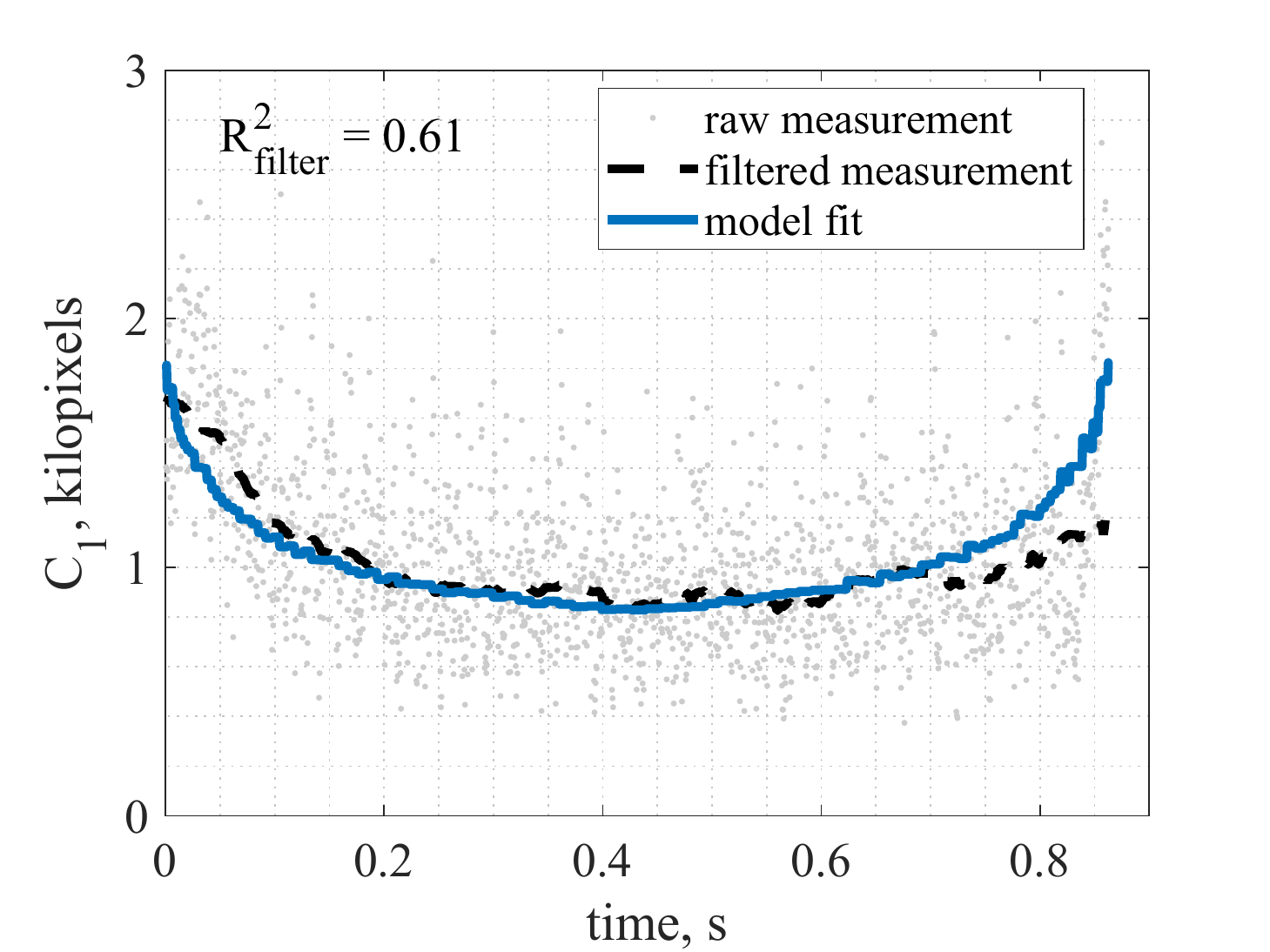}
    \caption{Illustration of the model fit on the validation data. One out of 24 validation layers shown. Notice the high variation of the raw measurement. However, general trend is captured well. To extract the trend, median filter $F$ with the window size of 150 samples was used.} 
    \label{fig:validation}
\end{figure}

Figure \ref{fig:validation} illustrates the validation results by showing a model fit on one of the validation layers from \textit{Cubes}. It appears that identified model captured the in-layer signal trend well: the $R^2$ on the filtered trend is $0.61$. With all 24 validation scans considered, the resultant $R^2 = 0.57$. As was described in Section \ref{sec:methods}, coaxial signal exhibits high variation. Therefore, filtered signal had to be used to quantify the fit to the geometry-related trend.
\FloatBarrier
\subsection{Control-oriented model.}
The meaning of the ``scan line length'' as a regressor should be clarified, as it is not obvious why the $C_1$ signature would depend on the line length. For the majority of part locations, there is a ``neighboring'' location on the build plate that was scanned in the previous scan line, by the nature of continuous laser scanning. As the laser is scanning the layer, heat left by the laser diffuses through the surroundings. Thus, if the ``neighboring'' location was visited not too long ago, the current point could see the effects of heat diffusion. If the length of the line currently being scanned is $l$, then, \textit{on average}, the time difference between the current point and its previously scanned ``neighbor'' is $\Delta t = l/V$, where $V$ is the laser scanning speed. Then, the model (\ref{eq:model}) can be interpreted as $C_1 \sim \exp(-\Delta t)$, which is reasonable from a cooling perspective. Moreover, as the scan length increases, $C_1$ approaches the limit defined by the laser power: the longer the line is, the more likely it is to sufficiently cool down by the time the laser comes back around.

Thus, the first model term $C_\infty(p)$ can be interpreted as a steady-state output corresponding to a specific power, with no dynamic effects related to the scan pattern. The second term is then the contribution of the \textit{previous} scanning to the $C_1$. Therefore, the exponential term should receive the values of laser power and scan line length from the \textit{previous} line, which yields 
\begin{equation}
    C_1(n) = C_\infty(p_n) + \Delta C(p_{n-1})e^{-\frac{l_{n-1}}{r(p_{n-1})}} \label{eq:model_final}
\end{equation}
where $n$ is the line index. This is the final model that was used for the model-based feedforward control in this work.
\FloatBarrier

\textit{Remark.} Strictly speaking, the ``revisiting'' time for a point depends on a point's location within a line. The scan patterns in this research are continuous and meandering, thus the point ``revisit'' time linearly varies from zero at a turnaround point to $2l/V$ at the other end of the line. This ``proper revisit'' time was investigated as a regressor, and it showed a trend similar to the one shown in Figure \ref{fig:expon_fit}, however noisier.  It appears that the high inherent variation of the $C_1$ signal, superimposed on the linearly varying  ``proper revisit'' time, obscures the exponential trend that is otherwise evident when the ``average revisit'' time $l/V$ is used.

\subsection{Model-based feedforward control}
Given the model (\ref{eq:model_final}), the problem of melt pool regulation can be formulated mathematically. Let $n=0,1,...,L$ be the index of a scan line; $l_n$ be the length of the scan line $n$; $p_n$ be the laser power commanded during the scanning of the line $n$. The laser power level during a scan of a line is constant. As such, all images acquired on that line are predicted to have the same value of $C_1 = C_1(n)$. Then, the control problem is formulated as the regulation of a varying coaxial signal $C_1(n)$, given by (\ref{eq:model_final}), to a constant reference value $C_{ref}$. It can be stated as an optimization problem: the goal is to find an optimal vector of line-by-line powers $\Tilde{P} \triangleq [p_0,p_1,...,p_L]$, such that the difference between all values of $C_1(n)$ and the reference $C_{ref}$ is minimized, considering all scan lines:
\begin{equation}
    \Tilde{P}^* = arg \min_{\Tilde{P}} \sum_{n=0}^{L}||C_1(n) - C_{ref}||^2
    \label{eq:problem_simple}
\end{equation}

The optimal line-by-line power profile $\Tilde{P}^*$ should then reduce or completely eliminate the geometry-dependent behavior of the coaxial signature $C_1$.

\begin{figure}[bth]
    \centering
    \includegraphics[width=\linewidth]{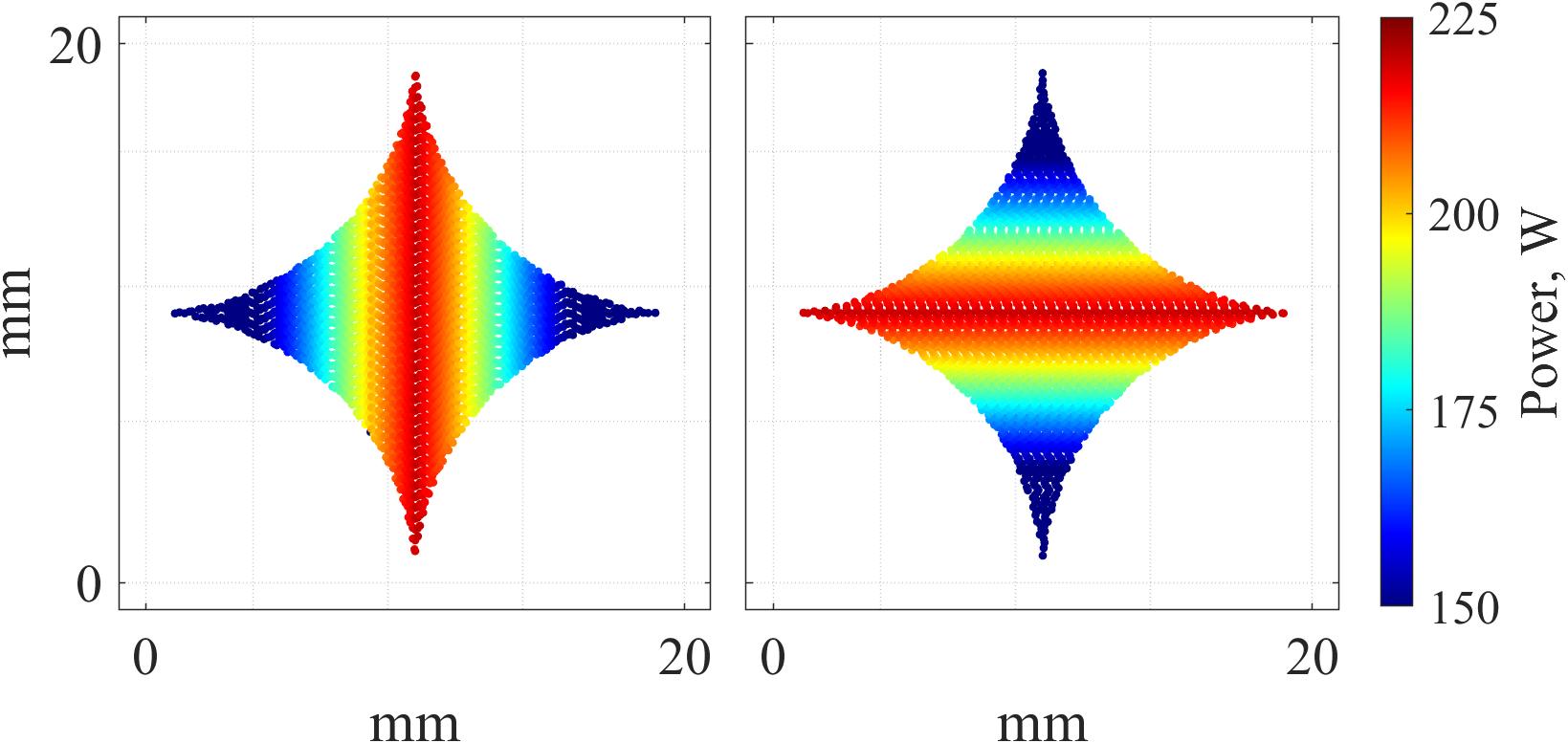}
    \caption{Optimal power profiles for \textit{Star} geometry. Horizontal and vertical scans are identical up to the rotation due to the part symmetry.}
   \label{fig:optimal_power_star}
\end{figure}

\begin{figure}[bth]
    \centering
    \includegraphics[width=\linewidth]{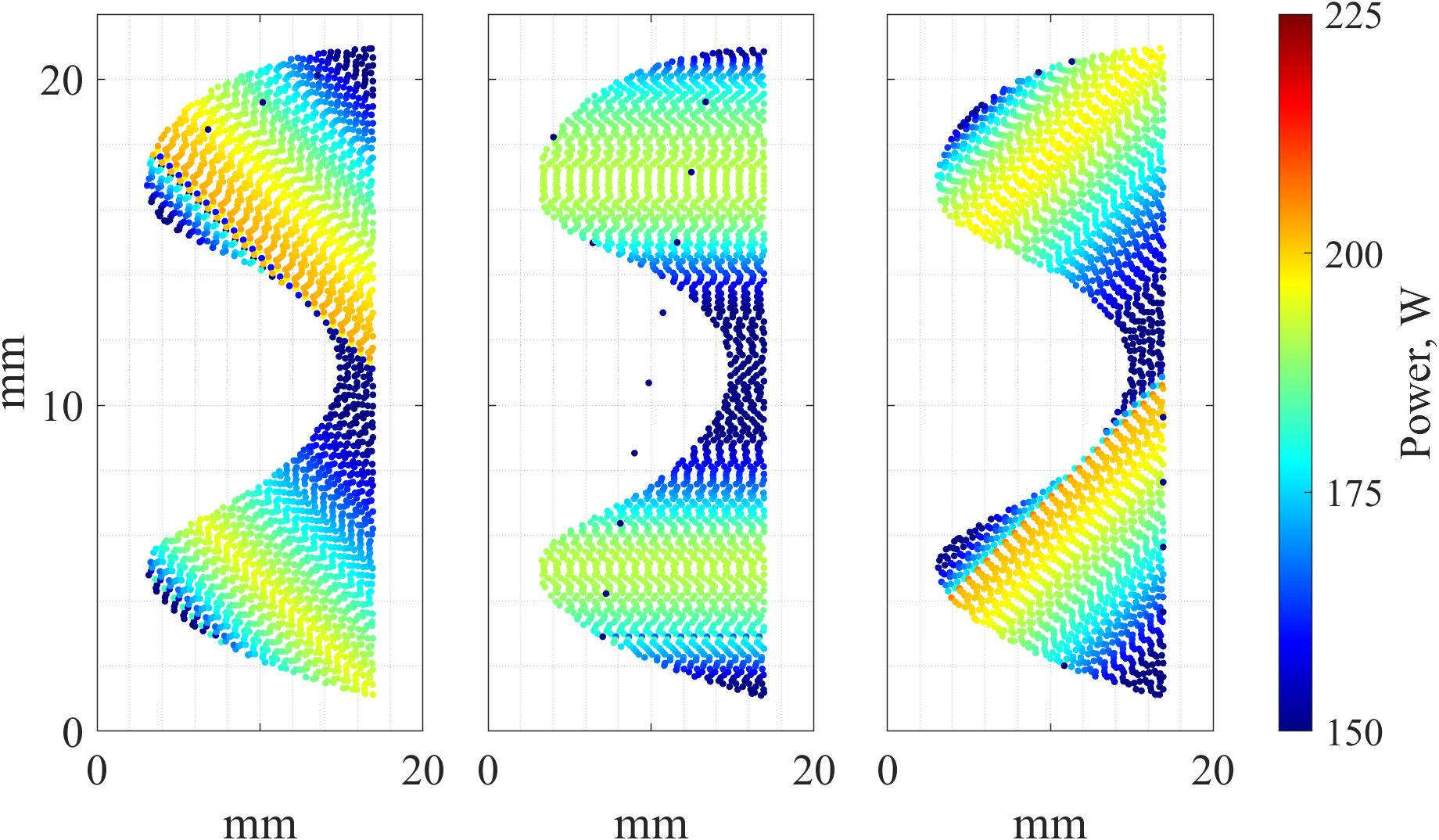}
    \caption{Optimal power profiles for \textit{Wave} geometry, 3 unique scanning angles.}
    \label{fig:optimal_power_wave}
\end{figure}

The optimal line-by-line power profiles were found for the \textit{Star} and \textit{Wave} parts, described in Section \ref{sec:methods}, via \texttt{MATLAB} with the \texttt{fmincon} function, for all unique scan patterns of these two geometries. These laser profiles are shown in Figures \ref{fig:optimal_power_star} and \ref{fig:optimal_power_wave}. The laser power was constrained to stay between 150 and 225 $W$, and the desired reference value was set to $C_{ref} = 1500$, which was the average value of $C_1$ for the nominal scan. Solving the optimization problem took approximately 8 seconds on a ThinkPad T495 laptop. 

\FloatBarrier

\begin{figure}[tbh]
    \centering
    \includegraphics[width=\linewidth]{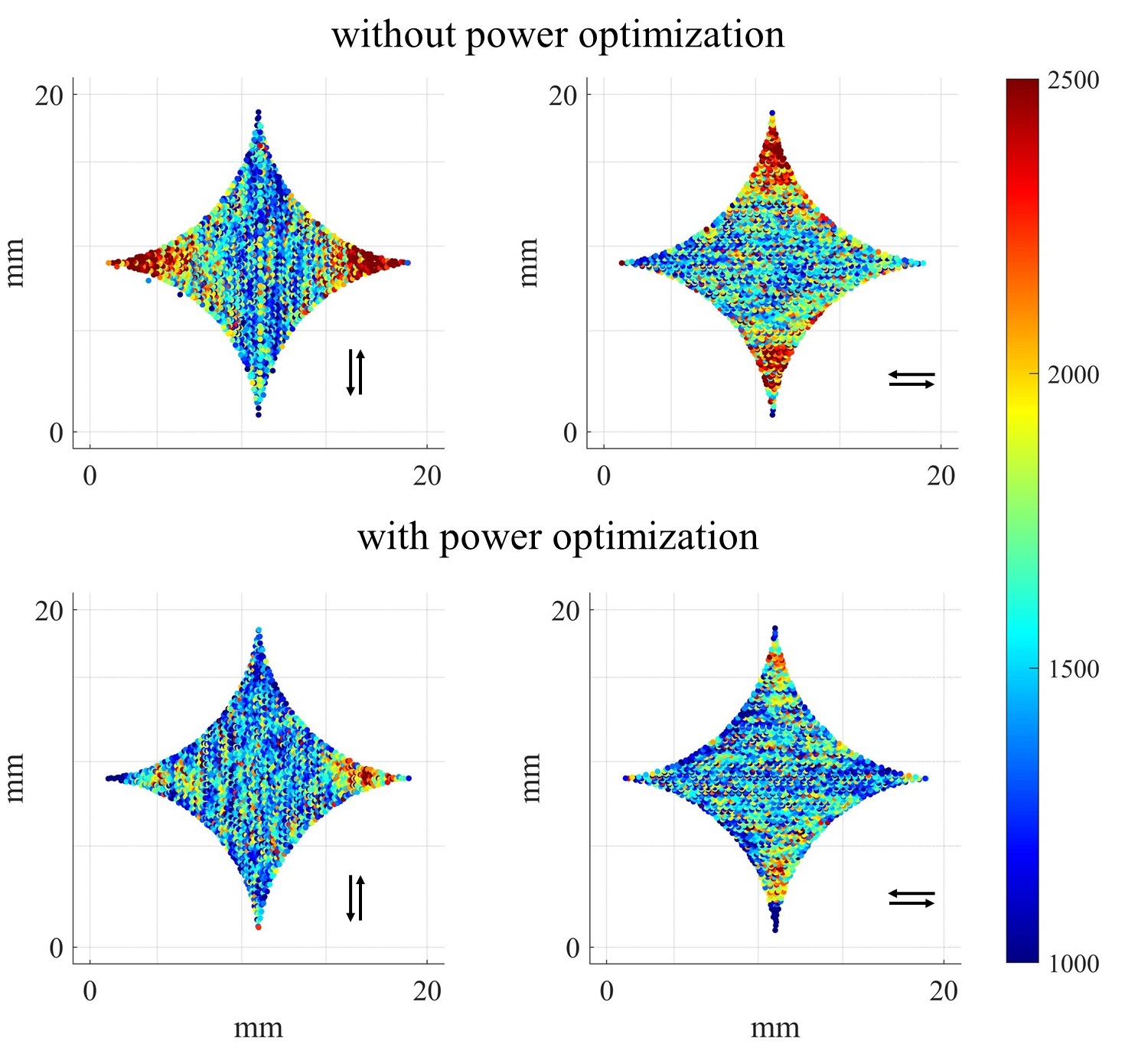}
    \caption{A comparison between $C_1$ coaxial signature with and without laser power optimization in \textit{Star} case. Laser scan direction is indicated by arrows. Representative layers shown. In-layer variation of $C_1$ is decreased with model-based laser power control.}
    \label{fig:ff_star}
\end{figure}

\begin{figure}[tbh]
    \centering
    \includegraphics[width=\linewidth]{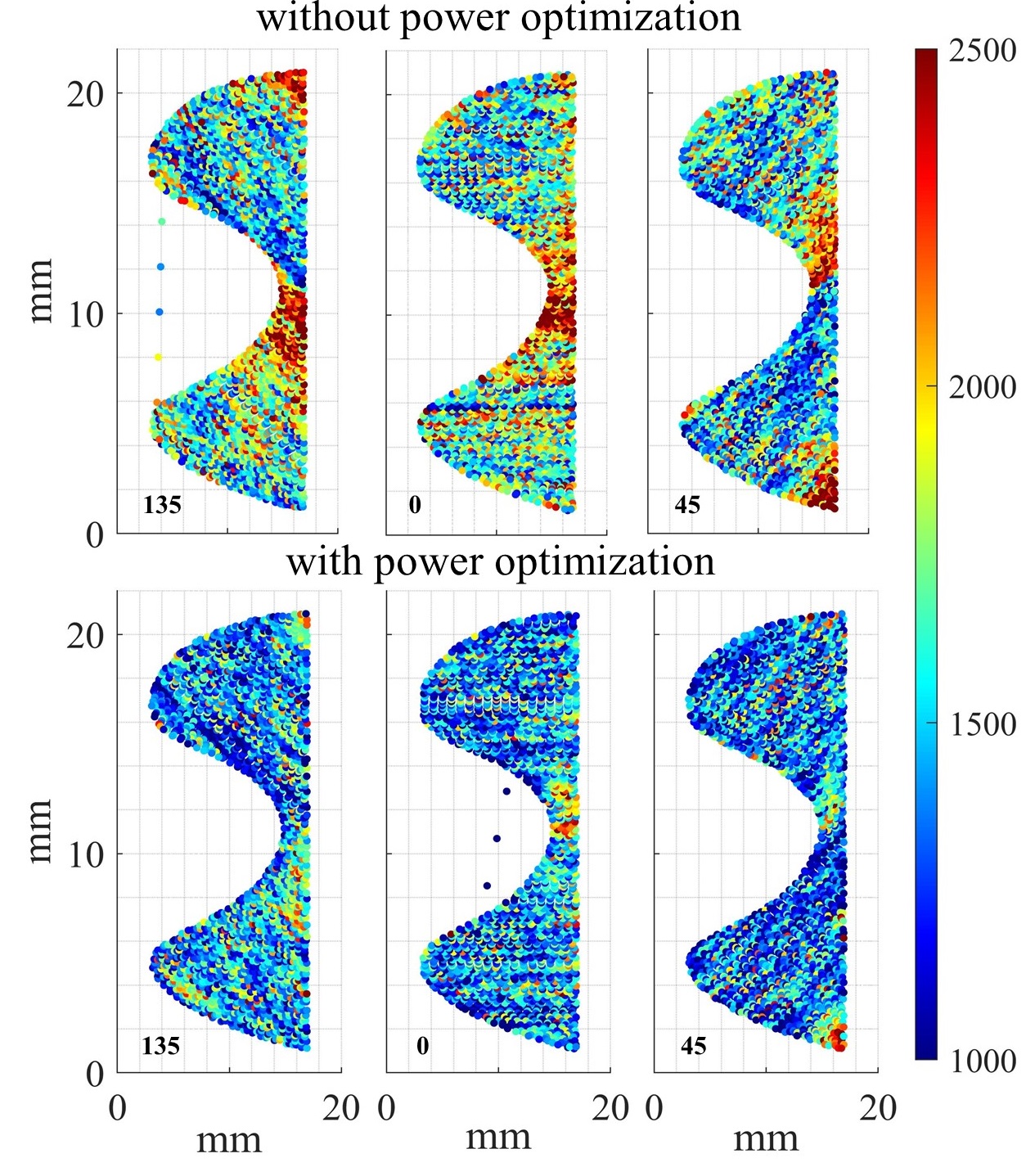}
    \caption{A comparison between $C_1$ coaxial signature with and without laser power optimization in \textit{Wave} case. Laser scan direction is indicated by angle value in degrees. Representative layers shown. In-layer variation of $C_1$ is decreased with model-based laser power control.}
    \label{fig:ff_wave}
\end{figure}

Resulting optimal power profiles were subsequently tested experimentally. For that, each of the layer scan files for the \textit{Star} and \textit{Wave} were modified such that each scan line was scanned with the appropriate power, as defined by the solution of the optimization problem (\ref{eq:problem_simple}). As evident from the resultant measurements, the optimal power profiles reduced the geometry-related effects. Figures \ref{fig:ff_star} and \ref{fig:ff_wave} compare $C_1$ signals between the controlled and the open-loop layers of the parts, showing that the geometry-dependent behavior of $C_1$ in corners and thin sections of the \textit{Star} and \textit{Wave} is visibly reduced in the controlled layers.

\FloatBarrier

\begin{figure}[tbh]
    \centering
    \subfigure[]{
    \includegraphics[width=0.47\linewidth]{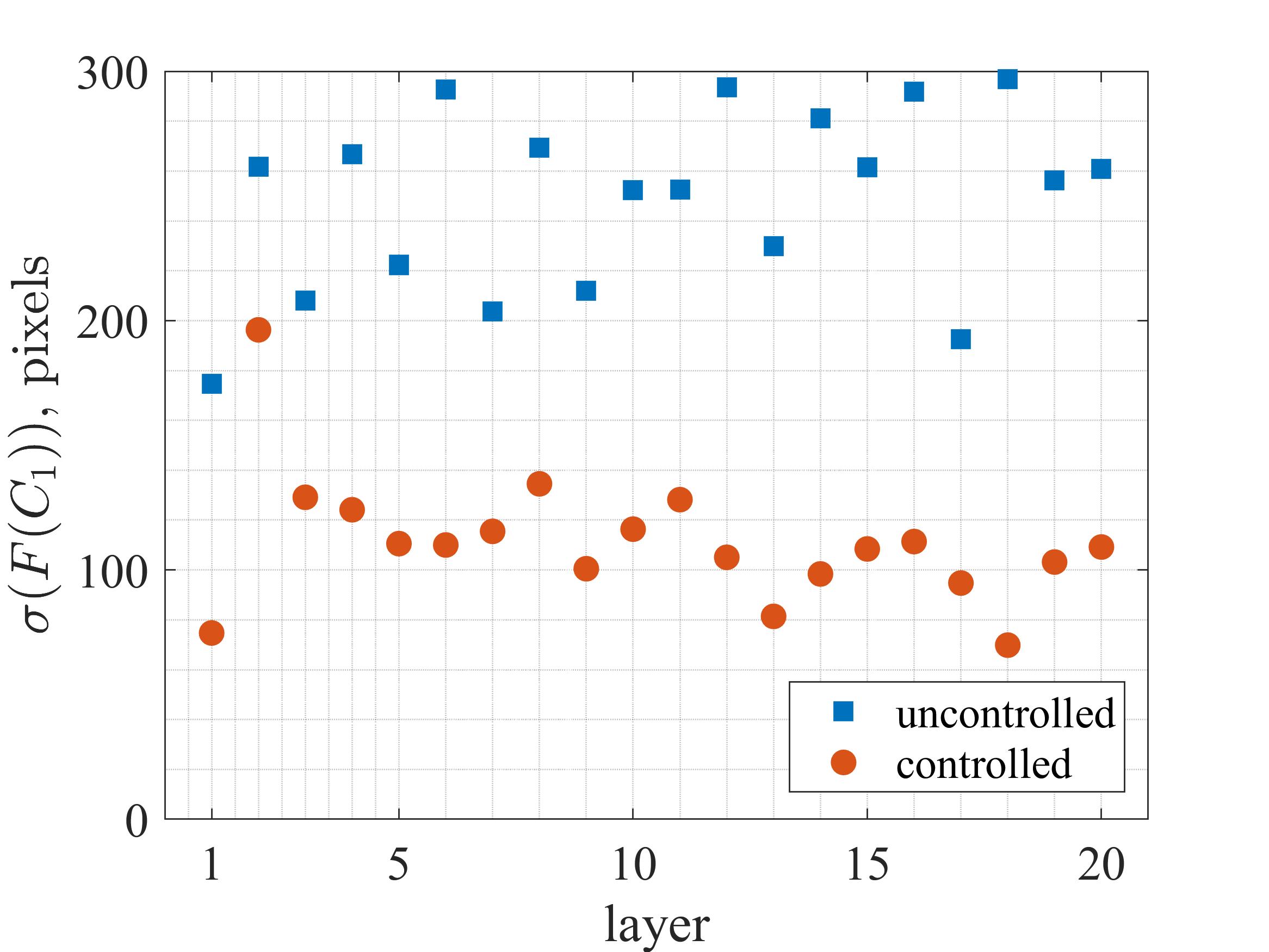}}
    \subfigure[]{
    \includegraphics[width=0.47\linewidth]{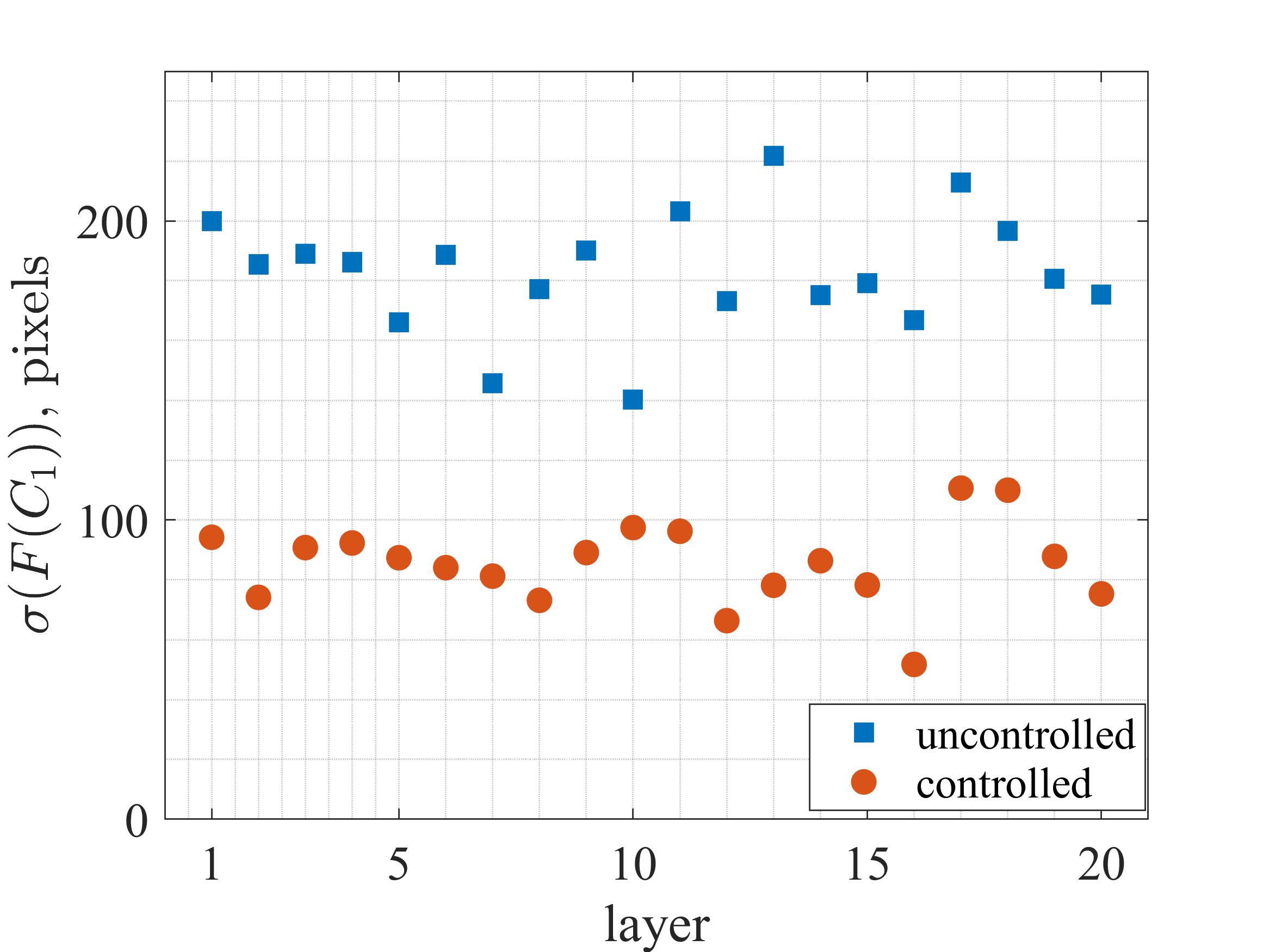}}
    \caption{A comparison between $C_1$ coaxial signature for the controlled and uncontrolled cases, for (a) \textit{Star} (b) \textit{Wave} parts. Signal variation within a layer is reduced approximately two-fold.}
    \label{fig:variance}
\end{figure}

As noted above, the $C_1$ signal has high variance. The same filter $F$ as in Figure \ref{fig:validation} can be used to quantify the trend, and once the filtered signal $F(C_1)$ is obtained, its variance $\sigma(F(C_1))$ can be used to quantify the geometry-related signal changes.  Figure \ref{fig:variance} shows the $\sigma(F(C_1))$ for 20 layers of the controlled \textit{Star} and \textit{Wave}, and compares it with the corresponding open-loop data. It is evident that, for both geometries, variation in the $C_1$ signal is reduced. For all 20 layers, the controlled part shows consistently lower values of $\sigma(F(C_1))$. Therefore, the proposed model-based controller reduced in-layer geometry-related variation in $C_1$ by around 50\%.

\section{Conclusion}
In this work, the behavior of the certain melt pool signature, i.e. an area of the melt pool ``footprint'', was shown to depend on the geometry of the layer scan pattern. It was demonstrated that the ``footprint'' exponentially increases in size in the areas of a part where scan lengths shorten, e.g. in corners or narrowing sections. The existence of such exponential behavior was observed on different part and scan geometries, including the ones with non-straight edges. To capture this behavior, the empirical model, incorporating variable laser power, was developed and experimentally validated. This model further enabled the application of a line-by-line model-based feedforward controller to regulate the ``footprint'' area. Given the empirical model, optimal laser power profiles to reduce the signal deviations were calculated and evaluated experimentally, for different geometries. The experimental evaluation of the proposed model-based feedforward control scheme demonstrated that such approach reduces the geometry-induced changes in ``footprint'' area two-fold, as compared to the open-loop operation. 

\section*{Acknowledgement}
This work was supported by NSF Data-Driven Cyberphysical Systems Award \#1645648 and by the State of New York ESD/NYSTAR program. 

\FloatBarrier

\bibliographystyle{ieeetr}
\bibliography{ref}

\begin{thebibliography}{10}

\bibitem{zhang2017defect}
B.~Zhang, Y.~Li, and Q.~Bai, ``Defect formation mechanisms in selective laser
  melting: a review,'' {\em Chinese Journal of Mechanical Engineering},
  vol.~30, no.~3, pp.~515--527, 2017.

\bibitem{wang2013research}
D.~Wang, Y.~Yang, Z.~Yi, and X.~Su, ``Research on the fabricating quality
  optimization of the overhanging surface in slm process,'' {\em The
  International Journal of Advanced Manufacturing Technology}, vol.~65,
  no.~9-12, pp.~1471--1484, 2013.

\bibitem{fox2016effect}
J.~C. Fox, S.~P. Moylan, and B.~M. Lane, ``Effect of process parameters on the
  surface roughness of overhanging structures in laser powder bed fusion
  additive manufacturing,'' {\em Procedia Cirp}, vol.~45, pp.~131--134, 2016.

\bibitem{King2015}
W.~E. King, A.~T. Anderson, R.~M. Ferencz, N.~E. Hodge, C.~Kamath, S.~A.
  Khairallah, and A.~M. Rubenchik, ``{Laser powder bed fusion additive
  manufacturing of metals: physics, computational, and materials challenges},''
  {\em Applied Physics Reviews}, vol.~2, no.~4, p.~041304, 2015.

\bibitem{kruth2007feedback}
J.-P. Kruth, P.~Mercelis, J.~Van~Vaerenbergh, and T.~Craeghs, ``Feedback
  control of selective laser melting,'' in {\em Proceedings of the 3rd
  international conference on advanced research in virtual and rapid
  prototyping}, pp.~521--527, Taylor \& Francis Ltd, 2007.

\bibitem{Renken2019}
V.~Renken, A.~V. Freyberg, K.~Sch{\"{u}}nemann, F.~Pastors, and A.~Fischer,
  ``{In‑process closed‑loop control for stabilising the melt pool
  temperature in selective laser melting},'' {\em Progress in Additive
  Manufacturing}, no.~0123456789, 2019.

\bibitem{Craeghs2011}
T.~Craeghs, S.~Clijsters, E.~Yasa, F.~Bechmann, S.~Berumen, and J.-P. Kruth,
  ``{Determination of geometrical factors in Layerwise Laser Melting using
  optical process monitoring},'' {\em Optics and Lasers in Engineering},
  vol.~49, no.~12, pp.~1440--1446, 2011.

\bibitem{Vasileska2020}
E.~Vasileska {\em et~al.}, ``{Layer-wise control of selective laser melting by
  means of inline melt pool area measurements},'' {\em Journal of Laser
  Applications}, vol.~32, no.~2, 2020.

\bibitem{Shkoruta2021}
A.~Shkoruta, S.~Mishra, and S.~J. Rock, ``{Real-Time Image-Based Feedback
  Control of Laser Powder Bed Fusion},'' {\em ASME Letters in Dynamic Systems
  and Control}, vol.~2, 07 2021.
\newblock 021001.

\bibitem{Shkoruta2019a}
A.~Shkoruta, W.~Caynoski, S.~Mishra, and S.~Rock, ``{Iterative learning control
  for power profile shaping in selective laser melting},'' in {\em IEEE
  International Conference on Automation Science and Engineering},
  vol.~2019-Augus, pp.~655--660, 2019.

\bibitem{Yeung2019}
H.~Yeung, B.~Lane, and J.~Fox, ``{Part geometry and conduction-based laser
  power control for powder bed fusion additive manufacturing},'' {\em Additive
  Manufacturing}, vol.~30, 2019.

\bibitem{Yeung2020-2}
H.~Yeung and B.~Lane, ``{A residual heat compensation based scan strategy for
  powder bed fusion additive manufacturing},'' {\em Manufacturing Letters},
  vol.~25, pp.~56--59, 2020.

\bibitem{Yeung2020-1}
H.~Yeung, Z.~Yang, and L.~Yan, ``{A meltpool prediction based scan strategy for
  powder bed fusion additive manufacturing},'' {\em Additive Manufacturing},
  vol.~35, 2020.

\bibitem{ren2021gaussian}
Y.~Ren and Q.~Wang, ``Gaussian-process based modeling and optimal control of
  melt-pool geometry in laser powder bed fusion,'' {\em Journal of Intelligent
  Manufacturing}, pp.~1--18, 2021.

\bibitem{Wang2020-penn}
Q.~Wang {\em et~al.}, ``{Model-based feedforward control of laser powder bed
  fusion additive manufacturing},'' {\em Additive Manufacturing}, vol.~31,
  2020.

\bibitem{AM_controller}
{America Makes}, ``{4039 Development \& demonstration of open-source protocols
  for powder bed fusion AM}.''

\bibitem{Grantham2016}
S.~Grantham, B.~Lane, J.~Neira, S.~Mekhontsev, M.~Vlasea, and L.~Hanssen,
  ``{Optical design and initial results from NIST's AMMT/TEMPS facility},''
  vol.~9738, p.~97380S, International Society for Optics and Photonics, may
  2016.

\bibitem{Demir2018}
A.~{G{\"{o}}khan Demir}, C.~{De Giorgi}, and B.~Previtali, ``{Design and
  implementation of a multisensor coaxial monitoring system with correction
  strategies for selective laser melting of a maraging steel},'' {\em Journal
  of Manufacturing Science and Engineering}, vol.~140, no.~4, 2018.

\bibitem{Pacher2019}
M.~Pacher, L.~Mazzoleni, L.~Caprio, A.~G. Demir, and B.~Previtali,
  ``{Estimation of melt pool size by complementary use of external illumination
  and process emission in coaxial monitoring of selective laser melting},''
  {\em Journal of Laser Applications}, vol.~31, no.~2, p.~022305, 2019.

\end{thebibliography}

\end{document}